\documentclass{article}

\PassOptionsToPackage{x11names}{xcolor}
\PassOptionsToPackage{numbers, compress, sort, square}{natbib}
\usepackage[preprint]{neurips_2026}
\usepackage{xcolor}

\usepackage{amsmath,amsfonts,bm}

\def\eqref#1{equation~\ref{#1}}

\def\1{\bm{1}}

\DeclareMathAlphabet{\mathsfit}{\encodingdefault}{\sfdefault}{m}{sl}
\SetMathAlphabet{\mathsfit}{bold}{\encodingdefault}{\sfdefault}{bx}{n}

\usepackage[utf8]{inputenc} %
\usepackage{lipsum} %

\usepackage[T1]{fontenc}

\usepackage{etoolbox}
\newbool{includeappendix}
\setbool{includeappendix}{true}

\ifdefined\isoverfull
\else
\fi

\newcommand{\eg}{e.g., }
\newcommand{\ie}{i.e., }

\newcommand{\agentsmd}{context file}
\newcommand{\agentsmds}{context files}
\newcommand{\Agentsmd}{Context file}
\newcommand{\Agentsmds}{Context files}
\newcommand{\planbench}{\textsc{CTXbench}}
\newcommand{\swebench}{\textsc{SWE-bench}}
\newcommand{\claudecode}{\textsc{Claude Code}}
\newcommand{\codex}{\textsc{Codex}}
\newcommand{\sonnet}{\textsc{Sonnet-4.5}}
\newcommand{\gptfive}{\textsc{GPT-5.2}}
\newcommand{\gptfivemini}{\textsc{GPT-5.1 mini}}
\newcommand{\qwencode}{\textsc{Qwen Code}}
\newcommand{\gptoss}{\textsc{GPT-OSS-120b}}
\newcommand{\qwen}{\textsc{Qwen3-30b-coder}}

\newcommand{\noplan}{\textsc{None}}
\newcommand{\autoplan}{\textsc{LLM}}
\newcommand{\humanplan}{\textsc{Dev}}

\usepackage{subcaption}

\definecolor{my-full-blue}{HTML}{1F77B4}

\definecolor{my-full-orange}{HTML}{FF7F0E}
\definecolor{my-extra-orange}{HTML}{7C4514}

\definecolor{my-full-green}{HTML}{2CA02C}

\definecolor{my-full-red}{HTML}{d62728}

\definecolor{my-full-purple}{HTML}{9467bd}

\definecolor{swebenchcolor}{HTML}{d7ecff} %
\definecolor{planbenchcolor}{HTML}{d5f5db} %

\colorlet{my-blue}{my-full-blue!30}
\colorlet{my-orange}{my-full-orange!30}
\colorlet{my-green}{my-full-green!90}
\colorlet{my-red}{my-full-red!90}
\colorlet{my-purple}{my-full-purple!30}

\definecolor{mygreen}{HTML}{B5E3B5}
\colorlet{myred}{LightPink1}
\colorlet{myblue}{SteelBlue2}
\definecolor{mylightblue}{HTML}{B0DBFF}
\colorlet{myorange}{DarkOrange1}

\definecolor{darkblue}{RGB}{34,49,63}
\definecolor{lightblue}{RGB}{238,244,249}
\definecolor{accentblue}{RGB}{88,139,202}
\definecolor{lightgreen}{RGB}{220,245,230}
\definecolor{lightred}{RGB}{250,225,225}
\definecolor{darkgreen}{RGB}{40,167,69}
\definecolor{darkred}{RGB}{220,53,69}
\definecolor{textgray}{RGB}{120,120,120}
\definecolor{lightgray}{RGB}{240,240,240}   %
\definecolor{accentorange}{RGB}{127,17,144} %
\definecolor{gray2}{HTML}{FCFCFC}

\usepackage{listings}

\usepackage{textcomp}

\usepackage{xcolor}

\usepackage[scaled=0.8]{beramono}

\definecolor{ckeyword}{HTML}{7F0055}
\definecolor{ccomment}{HTML}{3F7F5F}
\definecolor{cstring}{HTML}{2A0099}

\lstdefinestyle{numbers}{
	numbers=left,
	framexleftmargin=20pt,
	numberstyle=\tiny,
	firstnumber=auto,
	numbersep=1em,
	xleftmargin=2em
}

\lstdefinestyle{layout}{
	frame=none,
	captionpos=b,
}

\lstdefinestyle{comment-style}{
	morecomment=[l]//,
	morecomment=[s]{/*}{*/},
	commentstyle={\color{ccomment}\itshape},
}

\lstdefinestyle{string-style}{
	showstringspaces=false,%
}

\lstdefinestyle{keyword-style}{
	keywordstyle={\ttfamily\bfseries},
	morekeywords={
		function,
		constructor,
		int,
		bool,
		return,
		returns,
		uint
	},
	morekeywords = [2]{},
	keywordstyle = [2]{\text},
	sensitive=true,
}

\lstdefinestyle{input-encoding}{
	inputencoding=utf8,
	extendedchars=true,
	literate=
	{ℝ}{$\reals$}1%
	{→}{$\rightarrow$}1%
	{α}{$\alpha$}1%
	{β}{$\beta$}1%
	{λ}{$\lambda$}1%
	{θ}{$\theta$}1%
	{ϕ}{$\phi$}1%
}

\lstdefinestyle{escaping}{
	moredelim={**[is][\color{blue}]{\%}{\%}},
	moredelim={**[is][\color{pastelgreen}]{??}{??}},
	mathescape=true
}

\lstdefinestyle{default-style}{
	basicstyle=\fontencoding{T1}\ttfamily\footnotesize,
	style=numbers,
	style=layout,
	style=comment-style,
	style=string-style,
	style=keyword-style,
	style=input-encoding,
	style=escaping,
	tabsize=2,
	upquote=true
}

\lstdefinelanguage{BASIC}{
	language=C++,
	style=default-style
}[keywords,comments,strings]%

\lstdefinelanguage{JavaScript}{
morekeywords=[1]{break, continue, delete, else, for, function, if, in,
new, return, this, typeof, var, void, while, with, const, let},
morekeywords=[2]{false, null, true, boolean, number, undefined, string,
Array, Boolean, Date, Math, Number, String, Object},
morekeywords=[3]{eval, parseFloat, escape, unescape},
sensitive,
morecomment=[s]{/*}{*/},
morecomment=[l]//,
morecomment=[s]{/**}{*/}, %
morestring=[b]',
morestring=[b]"
}[keywords, comments, strings]
\definecolor{delim}{RGB}{20,105,176}
\definecolor{numb}{RGB}{106, 109, 32}
\definecolor{string}{rgb}{0.64,0.08,0.08}

\lstdefinelanguage{json}{
    showspaces=false,
    showtabs=false,
    breaklines=true,
    postbreak=\raisebox{0ex}[0ex][0ex]{\ensuremath{\color{gray}\hookrightarrow\space}},
    upquote=true,
    morestring=[b]",
    morecomment=[l]//,
    stringstyle=\color{string},
    literate=
     *{0}{{{\color{numb}0}}}{1}
      {1}{{{\color{numb}1}}}{1}
      {2}{{{\color{numb}2}}}{1}
      {3}{{{\color{numb}3}}}{1}
      {4}{{{\color{numb}4}}}{1}
      {5}{{{\color{numb}5}}}{1}
      {6}{{{\color{numb}6}}}{1}
      {7}{{{\color{numb}7}}}{1}
      {8}{{{\color{numb}8}}}{1}
      {9}{{{\color{numb}9}}}{1}
      {\{}{{{\color{delim}{\{}}}}{1}
      {\}}{{{\color{delim}{\}}}}}{1}
      {[}{{{\color{delim}{[}}}}{1}
      {]}{{{\color{delim}{]}}}}{1},
}
\definecolor{dkgreen}{rgb}{0,0.6,0}
\definecolor{dred}{rgb}{0.545,0,0}
\definecolor{dblue}{rgb}{0,0,0.545}
\definecolor{lgrey}{rgb}{0.9,0.9,0.9}
\definecolor{gray}{rgb}{0.4,0.4,0.4}
\definecolor{darkblue}{rgb}{0.0,0.0,0.6}
\lstdefinelanguage{cpp}{
      breaklines=true,               
      postbreak=\raisebox{0ex}[0ex][0ex]{\ensuremath{\color{gray}\hookrightarrow\space}},
      deletekeywords={...},          
      escapeinside={\%*}{*)},                  
      language=C++,                
      keywordstyle=\color{purple},  
      morekeywords={string,float}, 
      identifierstyle=\color{black},
      stringstyle=\color{blue},      
      showspaces=false,               
      showstringspaces=false,        
      showtabs=false,                
      tabsize=5,                     
    }

\usepackage[english]{babel}
\usepackage{xurl}
\usepackage[breaklinks=true]{hyperref}
\definecolor{darkpastelblue}{HTML}{0279AF}
\hypersetup{colorlinks,
      linkcolor=darkpastelblue,
      citecolor=darkpastelblue,
      urlcolor=darkpastelblue,
      filecolor=darkpastelblue}
\usepackage{algorithm}
\usepackage{algorithmicx}
\usepackage[noend]{algpseudocode}
\usepackage{color,soul}
\usepackage{ bbold }
\usepackage{multirow}
\usepackage{multicol}
\usepackage{caption}
\usepackage{tabularx,booktabs,xltabular}
\usepackage{graphicx}
\usepackage{tabu}
\usepackage{array}
\usepackage{siunitx}
\usepackage{subcaption}
\usepackage{fontawesome}
\usepackage{stackengine}
\usepackage{svg}
\usepackage{mathtools}
\usepackage{xspace}
\usepackage{wrapfig}
\usepackage{makecell}
\usepackage{placeins}
\usepackage{float}
\usepackage{pifont}
\usepackage{svg}
\usepackage[most]{tcolorbox}
\usepackage{array}
\usepackage{dsfont}
\usepackage{enumitem}
\usepackage[utf8]{inputenc}
\usepackage{textcomp}
\usepackage{amsthm}
\usepackage{amssymb,amsmath}
\usepackage[nounderscore]{syntax}
\usepackage{semantic}

\newcolumntype{x}[2]{S[table-format=#1.#2,table-auto-round]}

\usepackage{tikz}

\usetikzlibrary{arrows}
\usetikzlibrary{automata}
\usetikzlibrary{calc}
\usetikzlibrary{backgrounds}
\usetikzlibrary{decorations.markings}
\usetikzlibrary{decorations.pathmorphing}
\usetikzlibrary{decorations.pathreplacing}
\usetikzlibrary{fit}
\usetikzlibrary{patterns}
\usetikzlibrary{positioning}
\usetikzlibrary{shadows}
\usetikzlibrary{shapes}
\usetikzlibrary{shapes.geometric}
\usetikzlibrary{arrows.meta}
\usetikzlibrary{shadows.blur}

\definecolor{blue}{HTML}{347bc6}
\definecolor{green-underline}{HTML}{2de12c}
\definecolor{yellow-underline}{HTML}{ffd700}

\tikzstyle{block} = [
    minimum width=1cm,
    minimum height=0.5cm,
    align=center,
    fill=gray!30,
    rounded corners=5pt,
]

\tikzstyle{arrow} = [
	-{Latex[length=1.4mm, width=1.4mm]},
	draw=blue,
]

\tikzstyle{dashedline} = [
    draw=blue,
    dashed
]

\tikzstyle{line} = [
    draw=blue
]

\definecolor{lightblue}{RGB}{173,216,230} %

\usepackage{pgfplots}
\usepackage[capitalize,noabbrev]{cleveref}
\AtBeginDocument{
\crefname{appendix}{Appendix}{Appendices}
}

\usepackage{pifont}%

\definecolor{pastelgreen}{HTML}{059C05}
\definecolor{pastelred}{HTML}{FF7373}

\usepackage{setspace}

\newcommand{\ttt}[1]{\text{\texttt{#1}}}

\usepackage{scalerel}

\newcommand{\Hsquare}{%
  \text{\kern2\scriptspace\fboxsep=-.2pt\fbox{\rule{0pt}{1ex}\rule{1ex}{0pt}}\kern2\scriptspace}%
}

\algrenewcommand\alglinenumber[1]{\scriptsize #1\hspace{1mm}}
\newcounter{algoline}[algorithm]

\definecolor{PromptAccent}{HTML}{51a35f}

\newcommand{\placeholder}[1]{%
  \tcbox[on line,
    colback=PromptAccent!9,
    colframe=PromptAccent!35,
    boxrule=0.35pt,
    arc=0.9mm,
    left=2.5pt,right=2.5pt,top=1pt,bottom=1pt
  ]{\ttfamily\footnotesize\{#1\}}%
}

\newtcolorbox{promptbox}[1][]{%
  enhanced,
  breakable,
  colback=white,
  colframe=black!16,
  boxrule=0.55pt,
  arc=1.6mm,
  outer arc=1.6mm,
  left=10pt,right=10pt,
  top=8pt,bottom=8pt,
  borderline west={2.2pt}{0pt}{PromptAccent},
  drop shadow={black!12},
  fonttitle=\sffamily\bfseries\footnotesize,
  coltitle=PromptAccent!90!black,
  colbacktitle=PromptAccent!10,
  title={#1},
  attach boxed title to top left={xshift=10pt,yshift*=-2mm},
  boxed title style={
    enhanced,
    arc=1.2mm,
    boxrule=0.35pt,
    colframe=PromptAccent!35,
    colback=PromptAccent!10,
    left=6pt,right=6pt,top=2pt,bottom=2pt
  },
  before skip=10pt,
  after skip=10pt
}

\newtcblisting{schemabox}{%
  enhanced,
  breakable,
  colback=black!2,
  colframe=black!14,
  boxrule=0.4pt,
  arc=1.2mm,
  left=8pt,right=8pt,
  top=6pt,bottom=6pt,
  listing only,
  listing options={
    basicstyle=\ttfamily\footnotesize,
    breaklines=true,
    columns=fullflexible,
    keepspaces=true,
    showstringspaces=false
  }
}

\usepackage[capitalize]{cleveref}

\crefformat{section}{\S#2#1#3}

\crefrangeformat{section}{\S#3#1#4\crefrangeconjunction\S#5#2#6}

\crefmultiformat{section}{\S#2#1#3}{\crefpairconjunction\S#2#1#3}{\crefmiddleconjunction\S#2#1#3}{\creflastconjunction\S#2#1#3}

\newcommand{\crefrangeconjunction}{--}

\crefname{listing}{Lst.}{listings}
\crefname{line}{Line}{Lines}
\crefname{appendix}{App.}{App.}

\newcommand{\app}[1]{%
	\ifbool{includeappendix}{\cref{#1}}{the appendix}%
}
\newcommand{\App}[1]{%
	\ifbool{includeappendix}{\cref{#1}}{The appendix}%
}

\pgfplotsset{compat=1.18}

\title{Evaluating AGENTS.md:\\ Are Repository-Level Context Files Helpful for Coding Agents?}

\author{%
  Thibaud Gloaguen \\
  Department of Computer Science\\
  ETH Zurich\\
  \texttt{thibaud.gloaguen@inf.ethz.ch}
  \And
  Niels M{\"u}ndler-Sasahara \\
  Department of Computer Science\\
  ETH Zurich\\
  \texttt{niels.muendler@inf.ethz.ch}
  \And
  Mark Niklas Müller \\
  LogicStar.ai
  \And
  Veselin Raychev \\
  LogicStar.ai
  \And
  Martin Vechev \\
  Department of Computer Science\\
  ETH Zurich
}

\begin{document}

\maketitle

\begin{abstract}
A widespread practice in software development is to tailor coding agents to repositories using context files, such as \texttt{AGENTS.md}. Although this practice is strongly encouraged by agent developers, there is currently no rigorous investigation into whether such context files are actually effective for real-world tasks. In this work, we study this question and evaluate coding agents' task completion performance in two complementary settings: established SWE-bench tasks from popular repositories, with LLM-generated context files, and a novel collection of issues from repositories containing developer-committed context files.
Surprisingly, we find that providing context files does not generally improve task success rates, while \emph{increasing inference cost} by over 20\% on average. This observation holds across different LLMs, coding agents, and for both LLM-generated and developer-committed context files. Specifically, we find that while instructions in the context files are well followed by coding agents, repository overviews, although popular and recommended by model providers, are not helpful. We conclude that while context files are useful for specifying non-standard coding practices, any attempts to improve performance should be rigorously evaluated before deployment.
\end{abstract}

\vspace{-2mm}
\section{Introduction}
\label{sec:introduction}

Coding agents are being rapidly adopted across the software engineering industry \citep{aitw2025dashboard,cursor}, and providing \agentsmds{} like \texttt{AGENTS.md}, a \texttt{README} specifically targeting agents, has become common practice. With various industry leaders \citep{AGENTSMd2025,UsingClaudeMdFiles} recommending this approach to adapt their agents to specific repositories, \agentsmds{} are now supported by most popular agent frameworks, and included in over 60'000 open-source repositories, as reported by \citet{AGENTSMd2025}. 

These \agentsmds{} typically contain a repository overview and information on relevant developer tooling, aiming to help coding agents to navigate a given repository more efficiently, run build and test commands correctly, adhere to style guides and design patterns, and ultimately to solve tasks to the user's satisfaction more frequently.
To date, despite their widespread adoption, the impact of \agentsmds{} on the coding agent's ability to solve complex software engineering tasks has not been rigorously studied.
This is due to two key challenges: i) because of their recent introduction, \agentsmds{} are not available for instances of prior benchmarks, and ii) popular, well-known repositories, typically used to create such benchmarks, are not representative of most codebases. As a result, a rigorous evaluation of real-world \agentsmd{} use requires a new, complementary benchmark that contains only issues from less popular repositories with developer-committed \agentsmds{}.

\begin{figure*}[t]
    \centering
    \includegraphics[width=.9\linewidth]{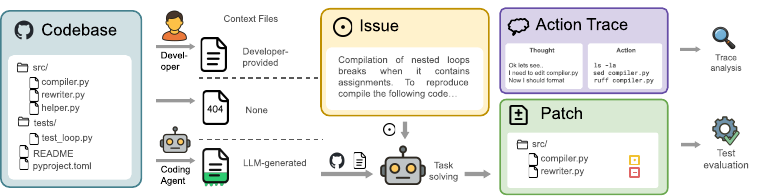}
    \caption{Overview of our evaluation pipeline. We begin with real-world repositories and tasks derived from past pull requests. For each repository state, we generate three settings: \textcircled{\tiny{1}} If a developer-provided \agentsmd{} exists, we include it in the repository. In \textcircled{\tiny{2}}, we omit the \agentsmd{}. \textcircled{\tiny{3}} We use the coding agent's recommended settings to generate the \agentsmd{}. Then we pass the repository and \agentsmd{} to the coding agent and instruct it to resolve the task. We finally analyze the trace for behavioral changes and apply the generated patch to check for task resolution success.}
    \label{fig:overview}
\end{figure*}

\paragraph{This work: Benchmarking \agentsmds{}' impact on resolving GitHub issues}
In this work, we investigate the effect of actively used \agentsmds{} on the resolution of real-world coding tasks.
We evaluate agents both in popular and less-known repositories, and, importantly, with \agentsmds{} committed by repository developers.
For this purpose, we construct a novel benchmark (\cref{fig:overview}, left), \planbench{}, comprising Python software engineering tasks, created specifically from real GitHub issues.
The benchmark contains $138$ unique instances, covering both bug-fixing and feature addition tasks across $12$ recent and niche repositories, all featuring developer-committed \agentsmds{}. \planbench{} complements \swebench{}, which we leverage for the evaluation of automatically generated \agentsmds{} on popular repositories.
We evaluate coding agents in three settings (\cref{fig:overview}, middle): without any \agentsmd{}, with \agentsmds{} automatically generated using agent-developer recommendations, and with the developer-committed \agentsmd{}.

We observe that including \agentsmds{} does not significantly affect performance. However, \emph{developer-committed} files outperform \emph{LLM-generated} ones by a significant margin of 7\% on average.
These observations are robust across LLMs, agents, and prompts used to generate the \agentsmds{}.
In a more detailed analysis of the agent traces, we observe that instructions in \agentsmds{} are well followed. This leads to increased exploration, testing, and reasoning by coding agents, and, as a result, increases costs by over 20\%.
We therefore suggest omitting LLM-generated \agentsmds{} for the time being, contrary to agent developers' recommendations, as they don't help improve performance.
Human-written \agentsmds{} should only include instructions required for coding agents that are not already present in the \texttt{README} (\eg specific conventions or non-functional requirements), and be rigorously evaluated before adoption.
We hope our evaluation framework will aid agent and model developers to improve the helpfulness of LLM-generated \agentsmds{}.

\paragraph{Key contributions} Our key contributions are:
\begin{enumerate}
    \item \planbench{}, a new benchmark for the impact of actively used \agentsmds{} on agents' ability to solve real-world software engineering tasks.
    \item An extensive evaluation of different coding agents on \planbench{} and \swebench{}, showing that \agentsmds{} don't significantly improve performance.
    \item A detailed investigation of agent traces, showing that \agentsmds{} instructions are well followed, leading to more thorough testing and exploration by coding agents.
\end{enumerate}

\section{Background and Related Work}
\label{sec:related-work}
\label{sec:background}

\paragraph{Coding agents} Coding agents are LLM-based systems designed for autonomous resolution of coding tasks \citep{yang2024sweagent}. Typically, they consist of a harness that allows an LLM to interact with its environment using specialized tools for, e.g., executing bash commands, conducting web searches, or reading, creating, or modifying files \citep{wang2025openhands, yang2024sweagent}.

Their impressive performance on repository-level coding tasks like SWE-bench \citep{swebench} led to rapid adoption in the software engineering community \citep{cursor} and the development of new agents by specialized companies \citep{AiderAI,wang2025openhands} and model providers \citep{openaiCodex2026,googleGeminiCli,QwenLMQwen3Coder,ClaudeCodeDocsOverview}. 
Model providers now train their LLMs to use the tools exposed by their harnesses~\citep{QwenLMQwen3Coder}, which can substantially improve coding ability relative to simpler harnesses~\citep{swebench-bash}.

\paragraph{\Agentsmds}
As coding agents were more broadly adopted, a need arose to provide them with additional context about the codebases they were working with \citep{WhyICreatedAgentsMd,builderioImproveYourAICodeOutputWithAgentsMd}.
To this end, model and agent developers recommend including \emph{\agentsmds{}}, such as \texttt{AGENTS.md} or \texttt{CLAUDE.md}, with codebases \citep{OpenAIAgenticAIFoundation2025,UsingClaudeMdFiles}.
Many agent harnesses even provide built-in commands to initialize such \agentsmds{} automatically using the coding agent itself, \eg{} by providing a dedicated \ttt{/init} command in the agent interface \citep{openaiCodex2026,QwenLMQwen3Coder,ClaudeCodeDocsOverview}.
At the time of writing, \citet{AGENTSMd2025} report that over 60'000 public GitHub repositories include a \agentsmd{}.

\paragraph{Evaluating \agentsmds{}}
Prior work collected and categorized the content of \agentsmds{} \citep{chatlatanagulchai2025agentreadmesempiricalstudy,mohsenimofidi2025contextengineeringaiagents}, deriving mostly descriptive metrics about their content without investigating their effectiveness \citep{githubHowToWriteAGreatAgentsMdLessons}.
While individual developers report anecdotal evidence of better alignment and solution capabilities when providing \agentsmds{} \citep{builderioImproveYourAICodeOutputWithAgentsMd,TheRiseOfAgentsMd2025}, we are the first to investigate the impact of actively used \agentsmds{} on agent behavior and performance at scale.

\paragraph{Repository-level evaluation}
Spearheaded by \citet{swebench}, evaluating coding agents on the autonomous resolution of real-world repository-level tasks quickly became the gold standard for assessing their capabilities. 
While initial work focuses on issue resolution \citep{swebench}, follow-up work proposed benchmarks on feature addition \citep{li-etal-2025-fea,du2025swedevevaluatingtrainingautonomous},
unit test generation \citep{mundler2024swt}, function generation \citep{liang2025languagemodelsreplaceprogrammers}, code performance \citep{he2025sweperflanguagemodelsoptimize}, and security \citep{chen2025secureagentbenchbenchmarkingsecurecode}.
Our work evaluates whether autonomous issue resolution and feature addition capabilities improve with actively used \agentsmds{}. 

Orthogonally, benchmarks have also been extended by mining more recent and more difficult problems \citep{badertdinov2025swerebenchautomatedpipelinetask,zhang2025swebenchgoeslive}, as well as instances focusing on end-user applications \citep{vergopoulos2025automatedbenchmarkgenerationrepositorylevel}.
We follow their approaches to mining novel task instances to obtain a set of tasks in repositories that feature \agentsmds{}.

\relax

\section{\planbench{}}
\label{sec:method}

In this section, we discuss the requirements for \planbench{}, a \textsc{SWE-Bench}-like benchmark that targets the evaluation of developer-provided \agentsmds{}, its generation process, and its statistics.

\subsection{Notation and Definitions}
\label{sec:notation-success}
We first introduce the notation to describe codebases, their test suites, and changes to these codebases in the form of patches.
Following the notation of \citet{mundler2024swt}, we denote a codebase, or repository, $R$ after applying patch $X$ as $R \circ X$.
Several patches can be applied sequentially, \ie{} $R \circ X \circ Y$ is the codebase $R$ after applying a first patch $X$ and then a second one $Y$.

\newcommand{\myexec}{\mathrm{exec}}
\newcommand{\dvS}{\mathcal{S}}
\newcommand{\dvT}{\mathcal{T}}
A \emph{test suite} $\dvT$ is a collection of tests that is used to validate the functionality of code in the repository.
Executing a test suite $\dvT$ on repository state $R$ returns $\myexec_R(\dvT) \in \{\textsc{pass}, \textsc{fail}\}$, indicating either that all tests in the suite passed or that at least one test failed.
An \emph{issue} $I$ is a task for autonomous completion by the coding agent, such as resolving a bug or implementing a requested feature.
We denote quadruples of $(I, R, \dvT, X^{*})$ as \emph{instances}, where the coding agent is tasked with predicting a patch $\hat{X}$ given issue $I$ and repository state $R$ such that $\myexec_{R \circ \hat{X}}(\dvT) = \textsc{pass}$, and $X^{*}$ is the golden patch for that instance.
We define the \emph{success rate} $\dvS$ as the percentage of predicted patches $\hat{X}_i$ for instances $(I_i, R_i, \dvT_i, X^{*}_i)$ where $\myexec_{R_i \circ \hat{X}_i}(\dvT_i) = \textsc{pass}$.

\subsection{Generation of \planbench{} Instances}
To construct \planbench{}, we use a five-stage construction process summarized below. 
We defer all the prompts used for this process to \cref{app:prompt}.

\paragraph{Requirements}
We aim to evaluate the impact of both automatically generated \agentsmds{} and developer-written \agentsmds{} on the success rate of coding agents on real-world tasks and codebases. 
The primary source for real-world codebases is open-source projects and their publicly tracked and documented changes, so-called pull requests (PRs). 
In order to obtain developer-written \agentsmds, we need to source PRs from projects that adopted \agentsmds{}. 
This is challenging, because \agentsmds{} were only formalized in August 2025, and have not been frequently used before.
Further, the adoption of \agentsmd{} is not uniform across the industry: even at the time of writing, many repositories do not include \agentsmds{}.

\paragraph{Finding repositories}
We first use GitHub search to build a list of potential candidate repositories to extract instances from.
Specifically, we select codebases that contain a \agentsmd{} such as \texttt{AGENTS.md} or \texttt{CLAUDE.md} at the root directory. Next, we filter down to those using Python as the main language and featuring a test suite. Finally, we filter for projects with many publicly documented changes, requiring at least $400$ PRs.
This criterion allows us to select codebases from which we can extract at least $10$ instances after our rigorous post-processing.

\paragraph{Filtering pull requests}
Given a repository, we filter PRs to retain those that are most likely to generate high-quality instances using a combination of rule-based checks and an LLM agent.
We only keep PRs that satisfy the following two criteria: they should reference at least one issue, and they should modify at least one Python file.
Further, we filter for PRs that are assessed by the agent to introduce deterministic, testable behaviors that are suitable for \swebench-like regression tests.
We notice that, because the use of \agentsmds{} is a recently emerging trend, most repositories containing \agentsmds{} are niche.
These niche repositories have less strict rules regarding pull requests, and thus most PRs may not include specific tests.
To enable building instances from these more niche repositories, we therefore do not require PRs to edit unit tests that validate the code changes but instead generate new ones (see below), in contrast to \swebench{}, which focused on large and popular repositories and requires PRs to contain unit tests.

\paragraph{Environment Set-Up}
For every PR and corresponding repository state, we set up an execution environment such that its test suite can be run, using a coding agent.
Specifically, we ask the agent to produce a small script that i) sets up the execution environment, ii) runs the test suite and iii) stores the results as a machine-readable dictionary at the root of the repository.
We only keep PRs where the resulting dictionary contains at least one passing test, which corresponds to $87\%$ of the filtered instances.

\paragraph{Task Descriptions}
Many of the smaller repositories we used to source \planbench{} do not enforce strict requirements on the quality of PR and issue descriptions. As a result, many issues are too imprecise and underspecified to solve the task in a testable manner (\eg in some cases, the PR body is empty).
Further, some PRs implement new features, which would require detailed descriptions about expected behavior and interfaces.
We therefore use a third LLM agent to produce a standardized and detailed task description $I$ based on the PR description, associated issues if available, and the original patch $X^{*}$.
This standardized task description is divided into 6 sections: description, steps to reproduce, expected behavior, observed behavior, specification, and additional information.
Importantly, we instruct the agent not to leak the solution in the generated task description, and to provide precise specifications. 
We inspected all generated instances, and found them to have high quality; none of them leaked the solution, whereas they showed sufficient specificity to correctly solve the given task.

\begin{wrapfigure}[13]{r}{0.47\textwidth}
    \vspace{-0.38in}
    \centering
    \includegraphics[width=\linewidth]{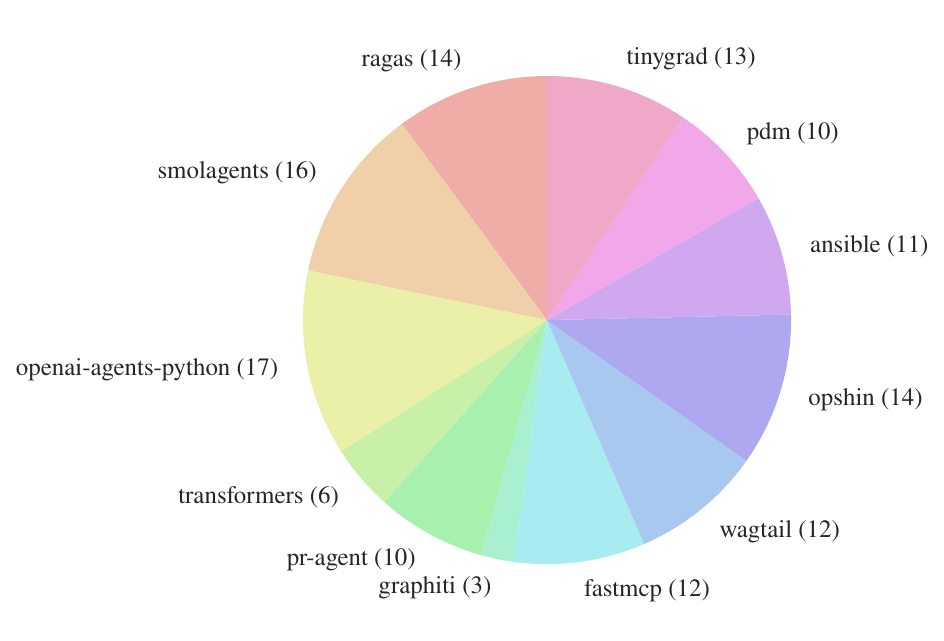}
    \caption{Distribution of \planbench{} instances across 12 open-source repositories.
    }
    \label{fig:overview_planbench}
    \vspace{-0.2in}
\end{wrapfigure}

\paragraph{Generating Unit Tests}
As most collected PRs do not modify or add unit tests that we could use to check the correctness of a given implementation, we use an LLM agent to generate such unit tests.
We provide the agent with the standardized task description $I$, the test files modified by the PR, if available, the original code changes $X^{*}$ made by the PR, and the base state of the repository $R$.
We then ask it to generate tests that pass for any implementation that resolves the described task. We verify that the added tests fail on $R$ and pass on $R \circ X^{*}$.
Finally, we manually review the generated tests and improve those overfitted to the reference implementation, resulting in newly generated tests $\mathcal{T}_i^{X}$.
We further determine all tests of the repository test suite $\mathcal{T}_i^{R}$ that pass on the patched code, i.e., the maximal set $\mathcal{T}_i^{R*} \subseteq \mathcal{T}_i^{R}$, such that $\myexec_{R_i \circ X_i^{*}}(\mathcal{T}_i^{R*}) = \textsc{pass}$, and obtain the final test set $\mathcal{T}_{i} = \mathcal{T}^{X}_{i} \uplus \mathcal{T}^{R*}_{i}$.
These tests achieve an average coverage of 75\% of the modified code (\cref{tab:overview_dataset}).
\vspace{-1mm}

\begin{wraptable}[14]{r}{0.5\textwidth}
    \vspace{-5mm}
\caption{Average, minimum, and maximum of key statistics of \planbench{} across the 138 instances.
For \agentsmds{}, a section is the content between Markdown headers.}
\vspace{-2mm}
\label{tab:overview_dataset}
\centering
\resizebox{\linewidth}{!}{
\begin{tabular}{ll ccc}
\toprule
&& Mean & Min & Max \\
\midrule
\multirow{1}{*}{PR body} & \# words & 415.3 & 5 & 4961 \\
\midrule
\multirow{1}{*}{Issue $I$} & \# words & 211.6 & 96 & 500 \\
\midrule
\multirow{1}{*}{Codebase} & \# files & 3337 & 151 & 26602 \\
\midrule 
\multirow{2}{*}{PR patch} & \# lines edited & 118.9 & 12 & 1973 \\
& \# files edited & 2.5 & 1 & 23 \\
\midrule 
\multirow{1}{*}{Test} & Coverage & 75\% & 2.5\% & 100\% \\
\midrule
\multirow{2}{*}{\Agentsmd} & \# words & 641.0 & 24 & 2003 \\
& \# sections & 9.7 & 1 & 29 \\
\bottomrule 
\end{tabular}}
\end{wraptable}

\paragraph{Evaluation}
We thus obtain \planbench{} instances $i$, each consisting of a task description $I_i$, a codebase $R_i$, golden patch $X^{*}_i$, and a set of tests $\mathcal{T}_{i}$.
During evaluation, we first set up the environment before prompting the coding agent with the task description $I_i$, retrieving the predicted patch $\hat{X}_i$, and measuring $\myexec_{R_i \circ \hat{X}_i}(\mathcal{T}_i)$.
\vspace{-1mm}

\paragraph{Overview of \planbench{}}
Using this process, we obtained 138 instances from a total of 5694 PRs from 12 repositories that meet our criteria, using \gptfive{} with \codex{} as the agent.
We visualize the distribution over repositories in \cref{fig:overview_planbench} and show key statistics of \planbench{} in \cref{tab:overview_dataset}. In comparison to \swebench{}, our dataset is both more evenly distributed over repositories and has otherwise similar statistics.
We validate the quality of the \planbench{} instances in \cref{app:analyzing_agentsmd}. Concretely, we find that the developer-provided \agentsmds{} from \planbench{} are significantly different from LLM-generated ones and thus likely to be human-written or edited. Additionally, we apply the LLM-aided task refinement to \swebench{}, showing that this process results in well-specified descriptions and tests, resulting in a fairer benchmark than the unrefined \swebench{}.
\vspace{-1mm}

\vspace{-1mm}
\section{Experimental Evaluation}
\label{sec:experiments}
\vspace{-1mm}

In this section, we investigate the effect of both LLM-generated and developer-provided \agentsmds{} on coding agent performance across models and agent harnesses.

\subsection{Experimental Setup}
\vspace{-1mm}

We describe the experimental setup below, deferring further details to \cref{app:experiments:setup}.
\vspace{-1mm}

\paragraph{Coding Agents}
We consider four coding agents, paired with suitable models: \claudecode{} \citep{ClaudeCodeDocsOverview} with \sonnet{}~\citep{sonnet45}, \codex{} \citep{openaiCodex2026} with \gptfive{} and \gptfivemini{} \citep{singh2025openaigpt5card}, and \qwencode{}~\citep{QwenLMQwen3Coder} with \qwen{}~\citep{qwen3}.
For \claudecode{}, we use the default settings and set the temperature of \sonnet{} to $0$.
Similarly, for \codex{}, we also use the default settings and set the temperature of \gptfive{} and \gptfivemini{} to $0$.
For \qwencode{}, we enable chat compression upon reaching $60$\% of the total context limit (set to $256$K tokens), restrict shell outputs to $2000$ tokens, and set the temperature of \qwen{} to $0.7$ with top-$p$ sampling at $0.8$. We deploy \qwen{} locally using vLLM~\citep{vllm}. We sample completions for each agent once.
For all agents, the \agentsmd{} is fed into their context, either by writing it to \texttt{AGENTS.md} for \codex{} and \qwencode{}, or to \texttt{CLAUDE.md} for \claudecode{}.
\vspace{-1mm}

\paragraph{Datasets}
We use the \textsc{Lite} split of \swebench{} \citep{swebench}, which consists of 300 tasks sourced from GitHub issues across 11 popular Python repositories, none containing developer-provided \agentsmds{}, and our novel \planbench{}, consisting of 138 instances from 12 repositories, all containing developer-provided \agentsmds{} (see \cref{sec:method}).
\vspace{-1mm}

\paragraph{Settings}
We consider three \agentsmd{} settings:
\par\vspace{-1mm}
\begin{tabular}[t]{l@{\hspace{2mm}}p{\dimexpr\linewidth-16mm\relax}@{}}
\noalign{}\vspace{1mm}
\noplan{} & No \agentsmd{} is available, i.e., we remove developer-provided files for \planbench{}. \\
\noalign{}\vspace{1mm}
\autoplan{} & An LLM-generated \agentsmd{} is available. We use the recommended initialization command and model for each agent individually to generate the \agentsmd{} using the pre-patch repository state $R$. \\
\humanplan{} & A developer-provided \agentsmd{} is available. We use the \agentsmd{} of the pre-patch repository state $R$. Only available for \planbench{}.
\end{tabular}

\paragraph{Metrics}
The main metric for agent performance is success rate (\cref{sec:notation-success}), i.e., the portion of instances for which the agent produces a patch that leads to all tests passing. 
We additionally consider the number of \emph{steps} the agent requires to complete a task. 
Each step is one interaction with the environment, e.g., calling a shell tool or modifying a file. 
Finally, we report the total \emph{cost} of LLM inference required to complete a task.
For \qwen{}, we estimate the cost from the average OpenRouter API price.
We report the statistical significance of changes in these metrics in \cref{tab:pvalue_resolution_rate,tab:cost_pvalue}.

\subsection{Main Results: Context Files Don't Improve Performance but Increase Cost}
\label{sec:eval:main}

\begin{figure*}[t]
    \centering
    \hspace*{\fill}
    \includegraphics[width=.48\linewidth]{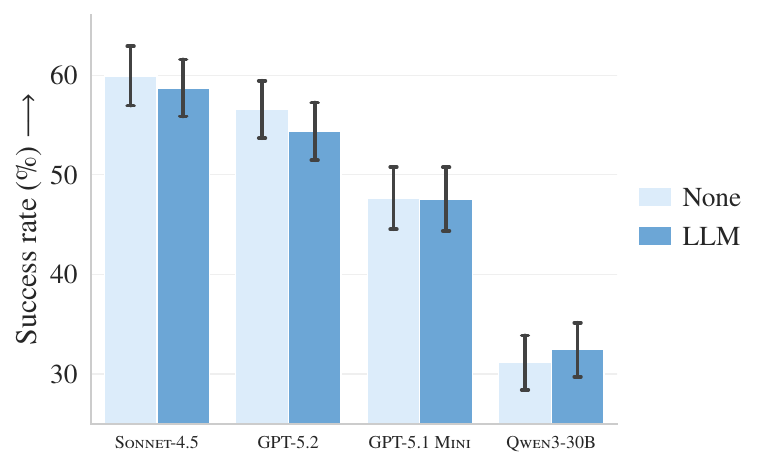}
    \hspace*{\fill}
    \includegraphics[width=.48\linewidth]{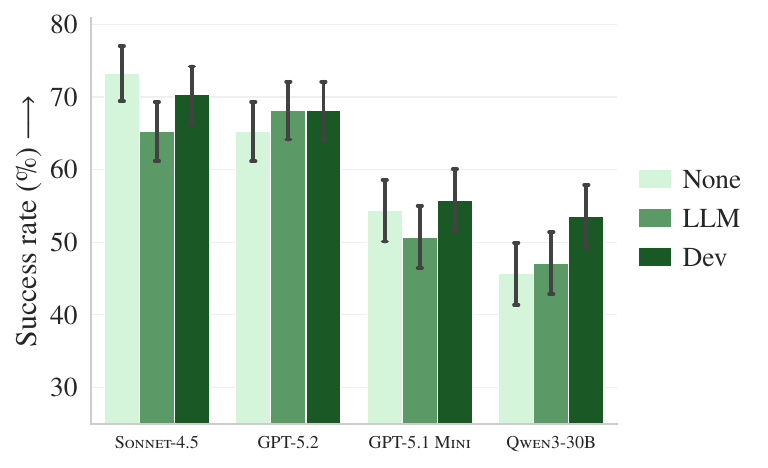}
    \hspace*{\fill}
    \caption{Resolution rate for 4 different models, without \agentsmds{}, with LLM-generated \agentsmds{}, and with developer-written \agentsmds{}, on \colorbox{swebenchcolor}{\swebench{}} (left) and \colorbox{planbenchcolor}{\planbench{}} (right).
    }
    \label{fig:experiment1_perf}
\end{figure*}

\begin{table}[t]
    \caption{The average number of steps (lower is better) and execution cost (in USD, lower is better) per SWE-Bench Lite and \planbench{} instance without \agentsmds{} (None), with LLM-generated \agentsmds{} (LLM), and with developer-\agentsmds{} (\textsc{Dev.}). We \textbf{bold} the best setting.}
    \label{tab:main_cost}
    \centering
    \resizebox{\linewidth}{!}{%
    \begin{tabular}{ll cc cc cc cc}
        \toprule
        & \multirow{2}{*}{Type} & \multicolumn{2}{c}{\textsc{Sonnet-4.5}} & \multicolumn{2}{c}{\textsc{GPT-5.2}} & \multicolumn{2}{c}{\textsc{GPT-5.1 M.}} & \multicolumn{2}{c}{\textsc{Qwen3-30B}} \\
        \cmidrule{3-10}
        && Steps & Cost & Steps & Cost & Steps & Cost & Steps & Cost \\
        \midrule
        \multirow[c]{2}{*}{\shortstack[c]{\textsc{SWE-}\\\textsc{Bench}}} & None & $\mathbf{54.4 \pm 2.2}$ & $\mathbf{1.30 \pm 0.07}$ & $\mathbf{12.5 \pm 0.8}$ & $\mathbf{0.32 \pm 0.05}$ & $\mathbf{40.9 \pm 4.2}$ & $\mathbf{0.18 \pm 0.03}$ & $\mathbf{29.7 \pm 1.9}$ & $\mathbf{0.12 \pm 0.01}$ \\
        \cmidrule{2-10}
        & LLM & $57.2 \pm 2.3$ & $1.51 \pm 0.08$ & $12.7 \pm 0.8$ & $0.43 \pm 0.05$ & $45.2 \pm 4.2$ & $0.22 \pm 0.03$ & $32.2 \pm 1.9$ & $0.13 \pm 0.01$ \\
        \midrule
        \multirow[c]{3}{*}{\shortstack[c]{\textsc{CTX-}\\\textsc{Bench}}} & None & $\mathbf{40.7 \pm 3.0}$ & $\mathbf{1.15 \pm 0.10}$ & $\mathbf{12.1 \pm 1.6}$ & $\mathbf{0.38 \pm 0.11}$ & $\mathbf{40.6 \pm 6.7}$ & $\mathbf{0.18 \pm 0.05}$ & $\mathbf{31.5 \pm 3.1}$ & $\mathbf{0.13 \pm 0.02}$ \\
        \cmidrule{2-10}
        & LLM & $46.5 \pm 3.9$ & $1.33 \pm 0.14$ & $13.1 \pm 1.5$ & $0.57 \pm 0.18$ & $46.9 \pm 6.4$ & $0.20 \pm 0.04$ & $34.2 \pm 3.4$ & $0.15 \pm 0.02$ \\
        \cmidrule{2-10}
        & \textsc{Dev.} & $45.3 \pm 3.6$ & $1.30 \pm 0.13$ & $13.6 \pm 1.7$ & $0.54 \pm 0.23$ & $46.6 \pm 7.1$ & $0.19 \pm 0.04$ & $32.8 \pm 3.7$ & $0.15 \pm 0.03$ \\
        \bottomrule
    \end{tabular}%
    }
    
\end{table}

\begin{wraptable}[12]{r}{0.5\textwidth}
    \vspace{-4mm}
    \caption{Two-sided Cochran-Mantel-Haenszel test comparing the effect of different context files on the resolution rate. The null hypothesis is that resolution rates are equal.
    We \textbf{bold} significant p-values.}\
    \label{tab:pvalue_resolution_rate}
    \centering
    \small
    \begin{tabular}{l l c}
        \toprule
        Benchmark & Comparison & p-value \\
        \midrule
        \swebench{}  & None vs LLM        & 0.87   \\
        \planbench{} & None vs LLM        & 0.37   \\
        \planbench{} & None vs Dev      & 0.21   \\
        \planbench{} & \textbf{LLM vs Dev}       & \textbf{0.038}  \\
        \bottomrule
    \end{tabular}
    \vspace{-0.1in}
\end{wraptable}

\paragraph{LLM-generated \agentsmds{} increase cost and don't improve performance}
LLM-generated \agentsmds{} cause performance drops in 5 out of 8 settings across \swebench{} and \planbench{} (see \cref{fig:experiment1_perf}).
Specifically, the average resolution rate is reduced by $0.5\%$ and $2\%$ on average on \swebench{} and \planbench{}, respectively.
With p-values of $87\%$ and $37\%$ (see \cref{tab:pvalue_resolution_rate}) under a two-sided test, this indicates that they have no significant effect on performance.
Meanwhile, they increase the \# steps in every setting, on average by $2.45$ and $3.92$, respectively, leading to a significant (p-value < 0.001\%) cost increase of 20\% and 23\% on average, respectively (see \cref{tab:main_cost}).

\paragraph{Developer-provided \agentsmds{} increase cost and outperform LLM-generated ones}
Developer-provided \agentsmds{} improve agent performance by $2.4\%$ on average ($p = 21\%$), significantly outperforming LLM-generated ones ($p=3.8\%$) (see \cref{tab:pvalue_resolution_rate}). Despite not being agent-specific, they improve performance for all agents but \claudecode{} (see \cref{fig:experiment1_perf} right).
However, developer-provided \agentsmds{} also increase the average number of steps and the cost, on average by $3.34$ steps and at most 19\%, respectively.

\subsection{Understanding Behavioral Changes Induced by Context Files}
\label{sec:eval:trace}

In this section, we investigate why \agentsmds{} do not affect performance meaningfully while increasing costs.
We find that, while instructions provided in \agentsmds{} are well followed, explaining the increase in cost, they do not provide effective repository overviews.

\begin{figure*}[t]
    \centering
    \hspace*{\fill}
    \includegraphics[width=.48\linewidth]{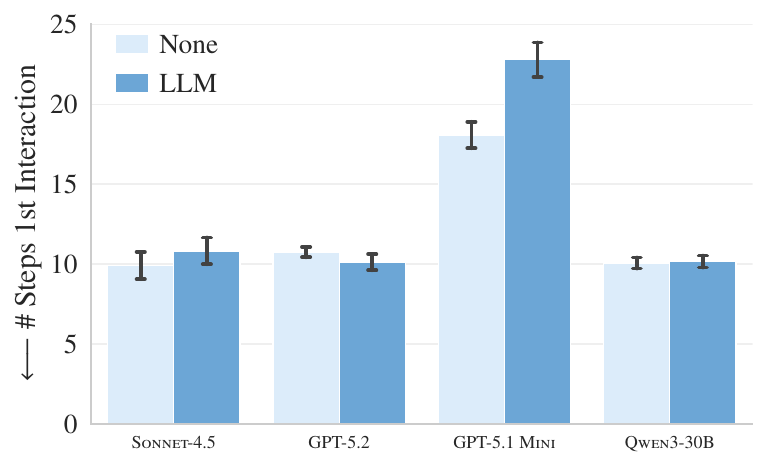}
    \hspace*{\fill}
    \includegraphics[width=.48\linewidth]{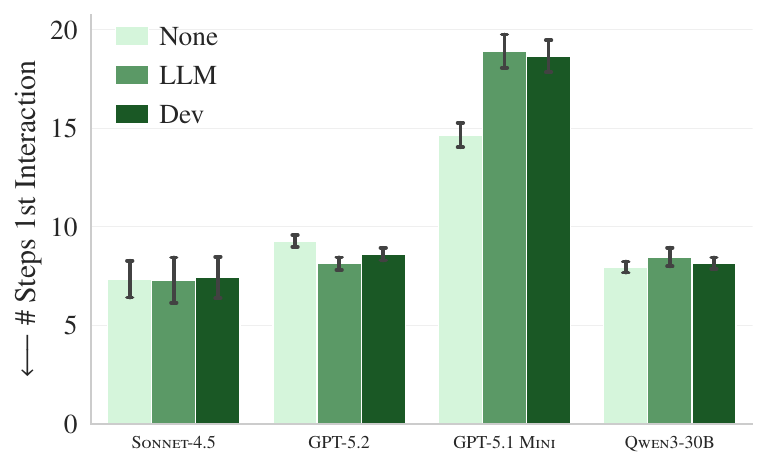}
    \hspace*{\fill}
    \caption{Number of steps before the first interaction between the agent and a file included in the PR patch (lower is better) is generally lower without \agentsmds{} than with LLM-generated \agentsmds{} or with developer-written \agentsmds{} (Human) on \colorbox{swebenchcolor}{\swebench{}} (left) and \colorbox{planbenchcolor}{\planbench{}} (right).
    }
    \label{fig:experiment1_speed}
\end{figure*}

\paragraph{\Agentsmds{} do not provide effective overviews}
One recommendation for \agentsmds{} is to include a codebase overview~\citep{AGENTSMd2025}.
Across the 12 developer-provided \agentsmds{} in \planbench{}, 8 include a dedicated codebase overview, with 4 explicitly enumerating and describing the directories and subdirectories in the repository.
Similarly, both the \codex{} and \qwencode{} \agentsmd{} generation prompts explicitly instruct the agent to include an overview section, while the \claudecode{} prompt advocates for a high-level overview only and warns against listing components that are easily discoverable.
We use \gptoss{} as a judge to assess which of the LLM-generated \agentsmds{} contain codebase overviews.
Surprisingly, $100\%$ of \sonnet{}-generated \agentsmds{} are flagged for overviews, and $95\%$ and $99\%$ are flagged for \qwen{} and \gptfive{}, respectively. Only \gptfivemini{} has significantly fewer overviews ($36\%$).

To assess the usefulness of these overviews, we measure how quickly agents discover files relevant to the described issue $I$.
In particular, we measure the average number of steps before the coding agent interacts with any file modified in the original PR patch $X^{*}$.
We exclude the $3\%$ of instances in which the agent never interacts with any file modified in $X^{*}$. 
Both on \swebench{} and \planbench{}, and for all agents, the \agentsmds{} do not meaningfully reduce this metric, while increasing the number of required steps significantly for \gptfivemini{}, as shown in \cref{fig:experiment1_speed}.

Inspecting the traces of \gptfivemini{} manually, we observe that the significant increase in the required number of steps is due to it (i) issuing multiple commands to find the \agentsmds{} and (ii) reading them (multiple times) despite them being already included in the agent's context. Interestingly, we only observed this behavior if \agentsmds{} were present at all.
We conclude that \agentsmds{} are not effective at providing a repository overview.

\paragraph{\Agentsmds{} lead to more testing, exploration, and specialized tool use}
Many \agentsmds{} contain recommendations on how to interact with the repository when solving issues.
To inspect whether agents follow these, we analyze the type and frequency of tool calls in agent traces.
For tools included in the agentic framework (\eg Read, Write, or Todo), we simply record the name of the tool being called.
For shell commands, we use an LLM-as-a-judge approach (using \textsc{GPT-OSS-120B}) to extract the executed commands (\eg \texttt{uv}, \texttt{pytest}, \texttt{cat}) and to categorize the intent of the tool call (\eg install dependencies, run tests, read files).
We build these categories iteratively, starting with an empty set, and allowing the judge to create new categories as needed (see \cref{app:experiments}).
\begin{figure*}[t]
    \centering
    \includegraphics[width=\linewidth]{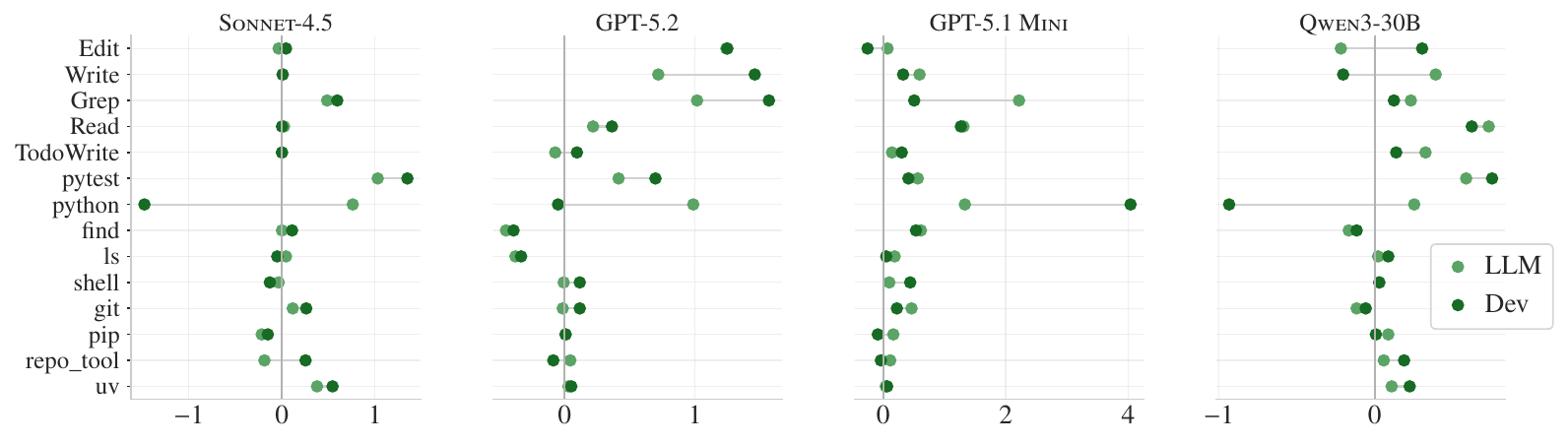}
    \caption{Increase in average tool use with LLM-generated (bright green) or developer-provided (dark green) \agentsmds{}, compared to the average tool use without \agentsmds{}. We map \codex{} and \qwencode{} tools to the \claudecode{} equivalents (we detail the mapping in \cref{app:experiments}).}
    \label{fig:experiment1_tool_freq}
    \vspace{-0.15in}
\end{figure*}

In \cref{fig:experiment1_tool_freq}, we show the increase in average tool use when including LLM-generated (bright green) or developer-provided (dark green) \agentsmds{}. Negative values imply a decrease in tool use. We find that, across all models, when \agentsmds{} are present, the coding agents run more tests.
They also tend to navigate the repository more: they search more files (\texttt{grep}), read more files, and write more files.
Lastly, adding \agentsmds{} causes agents to use more repository-specific tooling (\eg \texttt{uv} and \texttt{repo\_tool}).
In \cref{fig:experiment1_toolintent_freq} (\cref{app:experiments}), we perform a similar analysis using the intent of the tool call, and our conclusion is the same: \agentsmds{} lead to more testing and exploration.

\begin{wrapfigure}{r}{0.5\textwidth}
    \vspace{-0.2in}
    \centering
    \includegraphics[width=\linewidth]{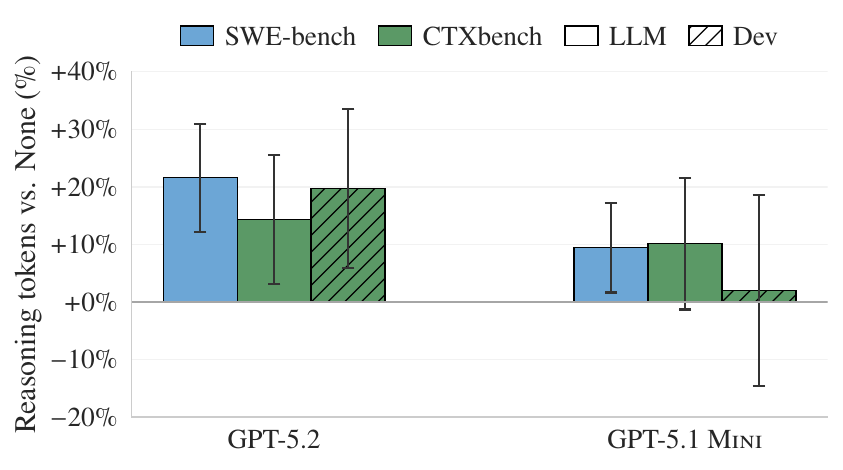}
    \caption{Relative change in average reasoning tokens spent by \gptfive{} and \gptfivemini{} compared to the no-\agentsmds{} baseline (None), computed on matched instances.}
    \label{fig:experiment1_ttc}
    \vspace{-0.3in}
\end{wrapfigure}

With respect to recommendations to use specific tools, we find that agents generally abide by them.
For instance, \texttt{uv} is used 1.6 times per instance on average when mentioned in the \agentsmds{}, compared to fewer than 0.01 times when it is not mentioned, and repository-specific tools are used 2.5 times per instance on average when mentioned, compared to fewer than 0.05 times when they are not mentioned.
In particular, this result implies that the absence of improvements when using \agentsmds{} is not due to a lack of instruction-following capabilities.
We defer a more in-depth analysis to \cref{app:prompts:trace_analysis}.

\paragraph{Following \agentsmds{} requires more thinking}
We hypothesize that these additional instructions make the task harder.
We analyze the average number of reasoning tokens used by \gptfive{} and \gptfivemini{}, as their adaptive reasoning~\citep{adaptive_reasoning} allows them to use more reasoning tokens for tasks that they deem harder.
\cref{fig:experiment1_ttc} shows the paired relative change compared to the no-\agentsmds{} baseline: LLM-generated \agentsmds{} increase the average number of reasoning tokens by 22\% for \gptfive{} and 10\% for \gptfivemini{} on \swebench{} (respectively 14\% and 10\% on \planbench{}), and developer-provided \agentsmds{} increase it by 20\% for \gptfive{} and 2\% for \gptfivemini{}.

\vspace{-0.1mm}
\subsection{Ablations}
\vspace{-0.1mm}

\begin{wrapfigure}[10]{r}{0.5\textwidth}
    \vspace{-5mm}
    \centering
    \includegraphics[width=.85\linewidth]{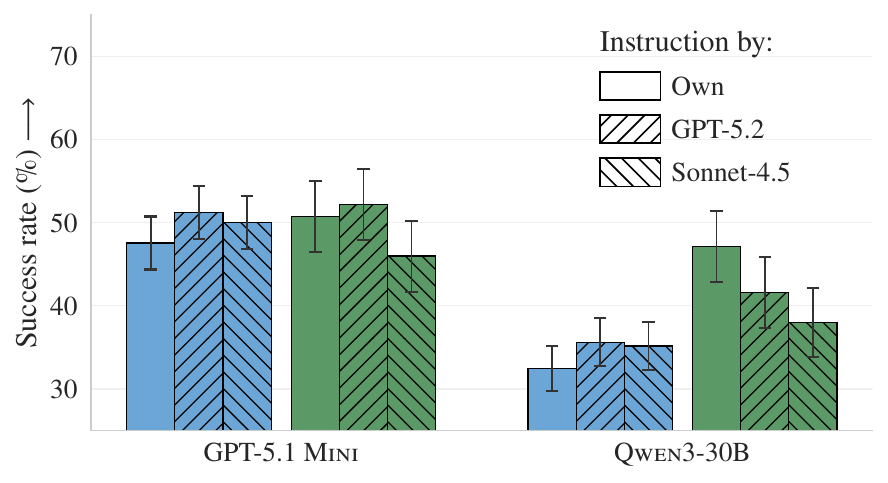}
    \caption{Comparison of \agentsmds{} effect across different generating models.
    }
    \label{fig:experiment2_perf}
\end{wrapfigure}

To investigate whether the poor performance of LLM-generated context files stems from a lack of model capabilities or bad prompts, we analyze differences between the \agentsmds{} generated by different models and the impact of the prompt used to create the \agentsmds{}.

\paragraph{Stronger models don't generate better \agentsmds{}}
We compare \agentsmds{} generated with \gptfive{} + \codex{} and \sonnet{} + \claudecode{} against self-generated ones when used by \gptfivemini{} and \qwen{}.
While performance improves on \swebench{} (2\% on average), it worsens on \planbench{} (3\% on average), showing that stronger models do not necessarily generate superior \agentsmds{}.

\begin{wrapfigure}[12]{r}{0.5\textwidth}
    \vspace{0mm}
    \centering
    \includegraphics[width=\linewidth]{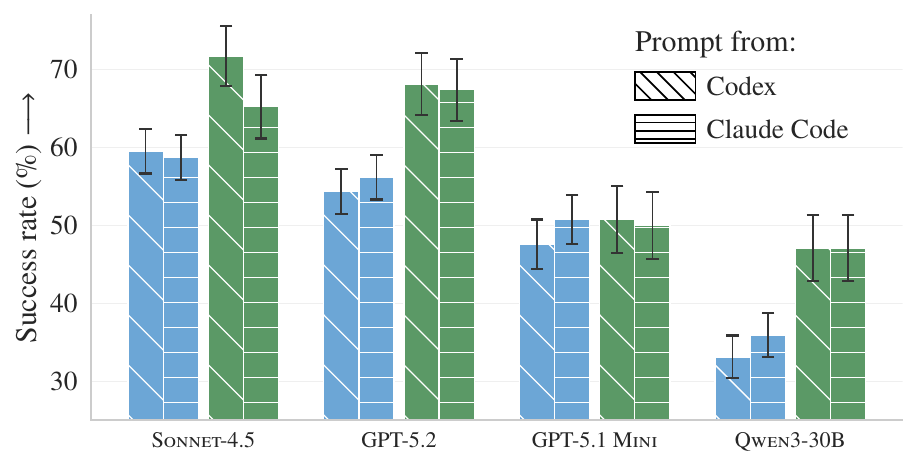}
    \caption{Ablation study on which prompt is used to generate \agentsmds{}.
    }
    \label{fig:experiment3_perf}
\end{wrapfigure}

\paragraph{No difference between the specific prompts}
We compare \agentsmds{} generated using the prompts of \codex{} and \claudecode{} across all agents and models in \cref{fig:experiment3_perf}.
Surprisingly, \claudecode{} performs better with \agentsmds{} generated using the \codex{} prompt, while both \gptfive{} and \gptfivemini{} perform better on \swebench{} with the \codex{} prompt but worse on \planbench{}.
Overall, neither the prompt that matches the underlying agent nor a specific prompt performs consistently best, indicating that sensitivity to good prompts is overall small.

\paragraph{Neither \agentsmd{} length nor specific instruction categories have a strong effect on performance}
In \cref{app:ablations}, we show that the length of \agentsmds{} has no significant impact on our results, that removing specific categories from LLM-generated \agentsmds{} does not change the results, and that contamination with task-specific knowledge cannot explain why \agentsmds{} do not improve performance.

\relax

\vspace{-0.1mm}
\section{Limitations and Future Work}
\label{sec:future-work}
\vspace{-0.1mm}

\paragraph{Programming languages}
The current evaluation is focused on Python. Since this is a language that is widely represented in the training data, detailed knowledge about tooling, dependencies, and other repository specifics might be present in the models' parametric knowledge, nullifying the effect of \agentsmds{}. Future work may investigate the effect of \agentsmds{} on more niche programming languages and toolchains that are less represented in the training data \citep{cassano2022multiplescalableextensibleapproach,pmlr-v202-orlanski23a}.

\paragraph{\Agentsmds{} beyond task resolution} In this work, we evaluate the impact of \agentsmds{} on task resolution rate. However, other aspects of coding agent performance, such as code efficiency \citep{he2025sweperflanguagemodelsoptimize} and security \citep{chen2025secureagentbenchbenchmarkingsecurecode}, would be interesting directions for future work. Security, specifically, was found to be sensitive to prompt changes in prior work \citep{baxbench}.

\paragraph{Improving \agentsmd{} generation} Another interesting avenue opened by this work is how to improve the automatic generation of \emph{useful} \agentsmds{}. Several related works in the direction of planning and continuous learning from prior tasks may be applicable to this task \citep{suzgun2025dynamiccheatsheettesttimelearning,zhang2025agenticcontextengineeringevolving,cheng2025evocurrselfevolvingcurriculumbehavior}. By tackling this challenge, future agents could gain a long-term capability at meaningful self-improvement.
Our work may serve as a baseline for how to rigorously evaluate automatically generated \agentsmds{}.

\section{Conclusion}
\label{sec:conclusion}

We evaluate the impact of \agentsmds{} on coding agent performance for four common coding agents on \swebench{} and the novel \planbench{}, built from recent GitHub issues and less popular repositories containing developer-written \agentsmds{}. 
We find that all \agentsmds{} consistently increase the cost and number of steps required to complete tasks. 
LLM-generated \agentsmds{} have a marginal negative effect on task success rates, while developer-written ones provide a marginal performance gain, neither statistically significant.

Our trace analyses show that instructions in \agentsmds{} are generally followed and lead to more testing and broader exploration; however, they do not function as effective repository overviews.
Overall, our results suggest that \agentsmds{} don't improve coding agent performance, and should only contain specific additional instructions beyond what is already available in the codebase.
This highlights a concrete gap between current agent-developer recommendations and observed outcomes, and motivates future work on principled ways to automatically generate concise, task-relevant guidance for coding agents.

\newpage
\bibliographystyle{plainnat}
\bibliography{references}

\clearpage
\appendix
\onecolumn

\section{Experimental Details}
\label{app:experiments}

In this section, we provide additional details about our experiments from \cref{sec:experiments}.

\subsection{Additional Experimental Details}
\label{app:experiments:setup}

We now describe the remaining experimental details for the experiments in \cref{sec:eval:main}.

\begin{figure*}[t]
    \centering
    \includegraphics[width=\linewidth]{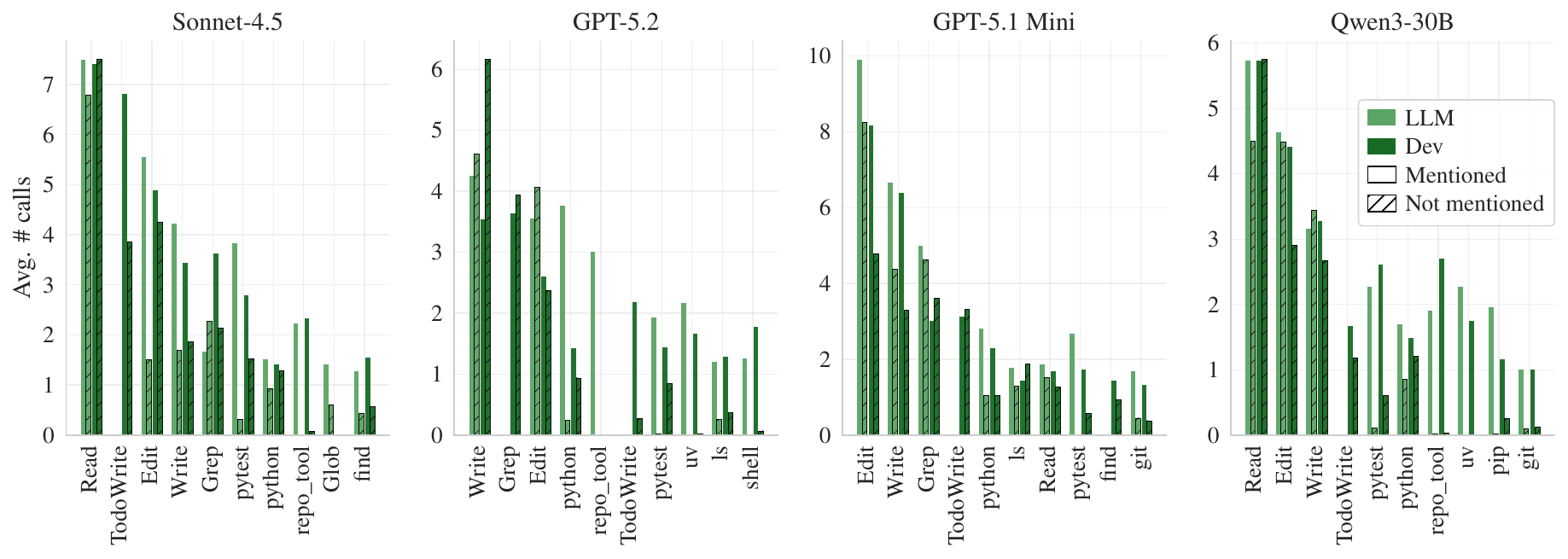}
    \caption{Average number of tool calls depending on whether the tool name is mentioned in the \agentsmds{}. For tool names, we use the equivalence classes from \cref{tab:mapping}, and consider a tool to be mentioned in the \agentsmd{} if any tool from the corresponding equivalence class is mentioned in the \agentsmd{}.}
    \label{fig:experiment1_toolcall_mention_correlation}
\end{figure*}
\vspace{-0.1in}

\paragraph{Coding environment}
For \planbench{} instances, we run the coding agent in a Docker container with basic tooling (python, apt-get, uv, \ldots) and Internet access.
Importantly, we remove the git commit history and all remotes.
For \swebench{}, we use the Docker images provided by~\citet{swebench}.
We let the coding agents access the web (either via dedicated tools or through the command line) and manually checked that the agents do not cheat (\eg looking at the PR corresponding to the instance description).
We find no such case of cheating, and web access represents a minority of the tool calls (less than 1\%).
For \claudecode{}, we keep the Task tool enabled: it allows \sonnet{} to invoke sub-agents, using \textsc{Haiku-4.5}, to solve sub-tasks.
For instance, a sub-task can involve exploring the repository to find specific files.

\subsection{Trace Analysis}

Here, we detail the experiments from \cref{sec:eval:trace}.
In particular, we give the mapping used to aggregate tool names across coding agents, analyze the correlation between the number of tool uses and whether the tool is mentioned in the \agentsmds{}, and expand the trace analysis to the intent behind tool calls.

\begin{wraptable}[10]{r}{0.5\textwidth}
  \centering
  \caption{Equivalence classes used to group the different tool calls.}
  \label{tab:mapping}
  \resizebox{\linewidth}{!}{%
    \begin{tabular}{lll}
      \toprule
      \claudecode{} tool & \codex{} & \qwencode{} \\
      \midrule
      Edit & \texttt{sed} & \texttt{sed}, \texttt{edit}  \\
      \midrule
      Write & \texttt{apply\_patch} & \texttt{write\_file}\\
      \midrule
      Grep & \texttt{grep}, \texttt{rg} & \texttt{grep} \\
      \midrule
      Read & \texttt{cat} & \texttt{cat}, \texttt{read\_file}, \texttt{search\_file\_content} \\
      \midrule
      TodoWrite & \texttt{update\_plan} & \texttt{todo\_write} \\

      \bottomrule
    \end{tabular}%
  }
\end{wraptable}

\paragraph{Experimental setup}
We recall the experimental setup from \cref{sec:eval:trace}.
Given a list of tool calls from an agent, we analyze the frequency of each tool call.
For tools included in the agentic framework (\eg Read, Write, or Todo), we simply record the name of the tool being called.
For shell commands, we use an LLM-as-a-judge to extract (from the command and its output) the concrete command that was executed (\eg \texttt{uv}, \texttt{pytest}, \texttt{cat}) and to categorize the intent of the tool call (\eg install dependencies, run tests, read files).
We build the categories iteratively.
We start with an empty set of categories, and for each shell command, we ask the judge to assign it to an existing category if possible and otherwise create a new category.
As the judge, we use \gptoss, and the prompt is given in \cref{app:prompts:trace_analysis}.
Finally, we manually merge duplicate and closely related categories.

\paragraph{Refining the tool names}
For \cref{fig:experiment1_tool_freq}, we further manually refined the tool names for readability.
In particular, in \cref{tab:mapping}, we map tool names from other agents, namely \codex{} and \qwencode{}, as well as some CLI tools, to \claudecode{} tooling.
Lastly, for repository-specific tooling (\eg \texttt{pdm}, \texttt{ansible}, or \texttt{opshin}), we grouped them into the \texttt{repo\_tool} category. \\

\begin{figure*}[t]
    \centering
    \hspace*{\fill}
    \includegraphics[width=0.97\linewidth]{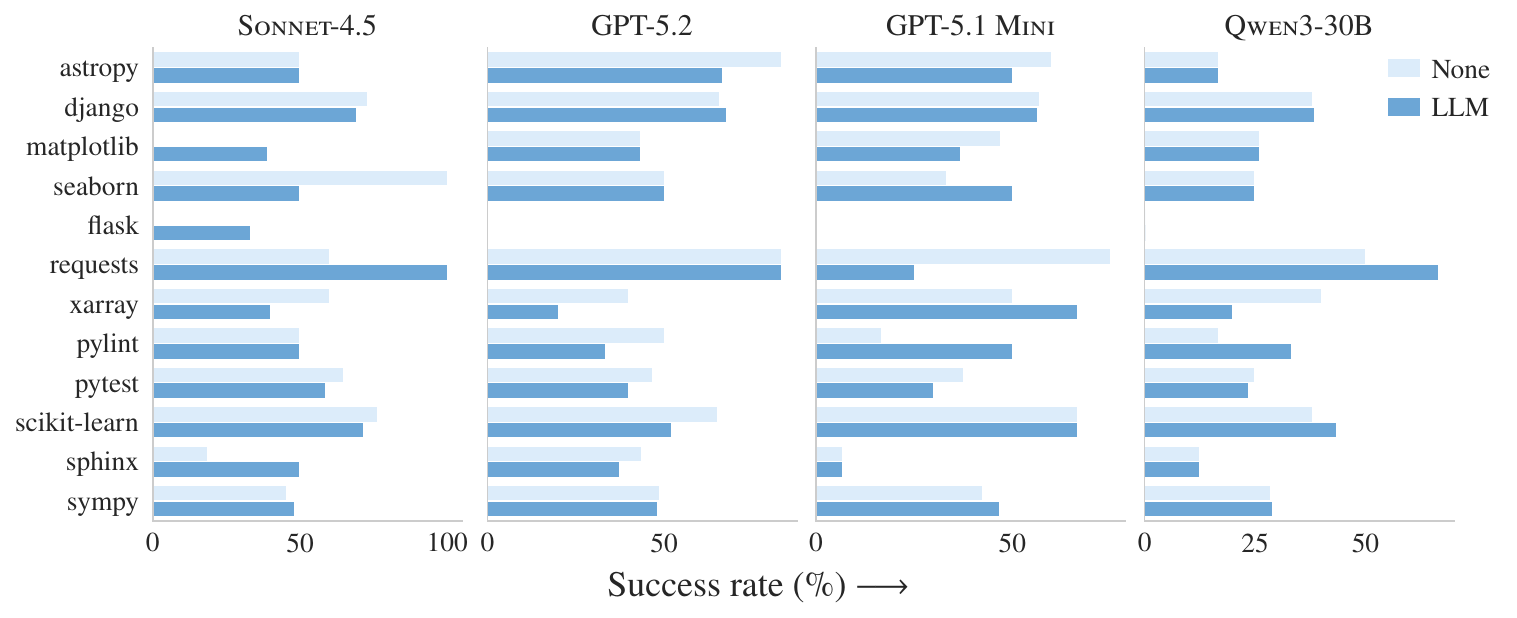}
    \hspace*{\fill}
    \includegraphics[width=0.97\linewidth]{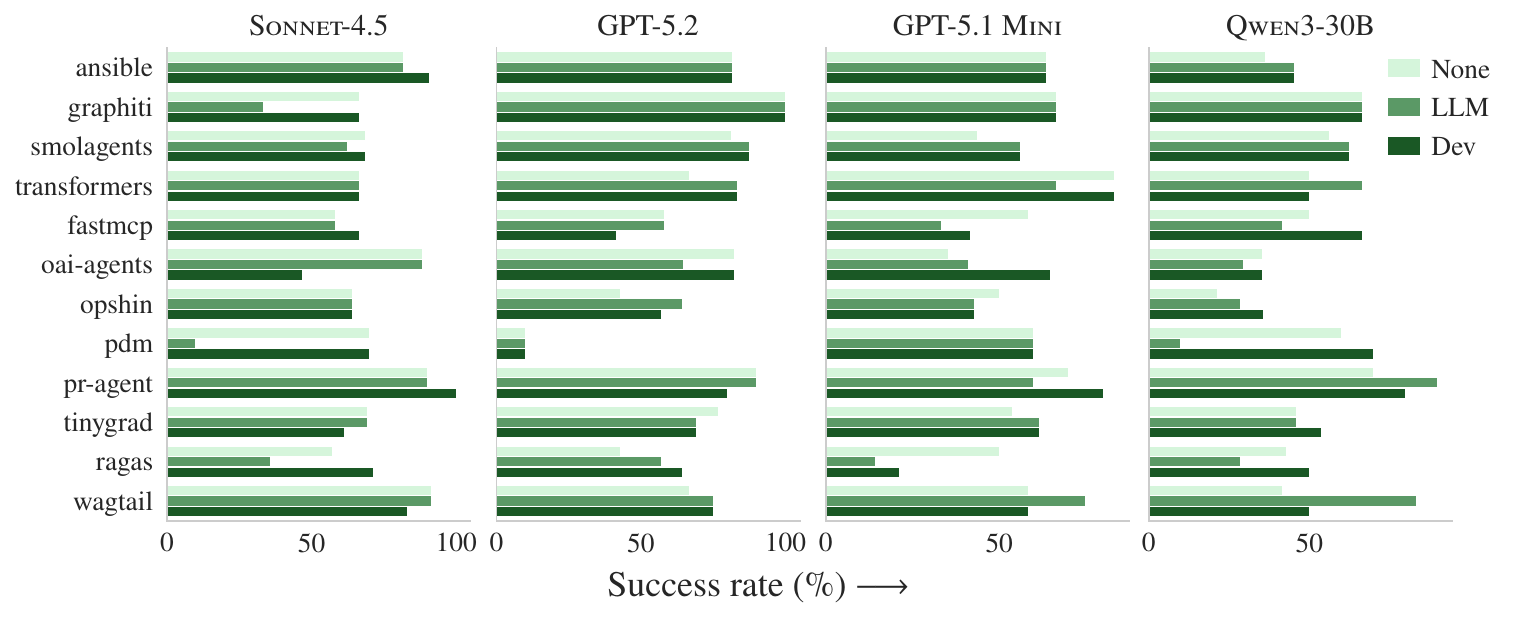}
    \hspace*{\fill}
    \caption{Resolution rate grouped by repository for four different models: without \agentsmds{}, with LLM-generated \agentsmds{}, and with developer-written \agentsmds{} on \colorbox{swebenchcolor}{\swebench{}} (top) and \colorbox{planbenchcolor}{\planbench{}} (bottom).
    For \swebench{} in particular, the majority of instances come from the same repository (\texttt{django}), making per-repository estimates of the success rate noisy.
    }
    \label{fig:experiment1_perf_pre_repo}
    \vspace{-0.15in}
\end{figure*}

\paragraph{Correlation between the number of tool calls and \agentsmds{}}
In \cref{fig:experiment1_toolcall_mention_correlation}, we show the average number of tool calls depending on whether the tool name is mentioned in the \agentsmd{}.
We find that, if a tool name is mentioned in the \agentsmds{}, this increases its usage by the coding agents.
For instance, \texttt{uv}, \texttt{pytest}, or repository-specific tools (\texttt{repo\_tool}) are used almost exclusively if they are mentioned in the \agentsmd{}.
This means that instructions in the \agentsmds{} are followed, and that a lack of instruction-following capabilities does not explain why we observe, in \cref{sec:eval:main}, no gain in accuracy when using \agentsmds{}.

\WFclear
\begin{figure*}[t]
    \centering
    \includegraphics[width=\linewidth]{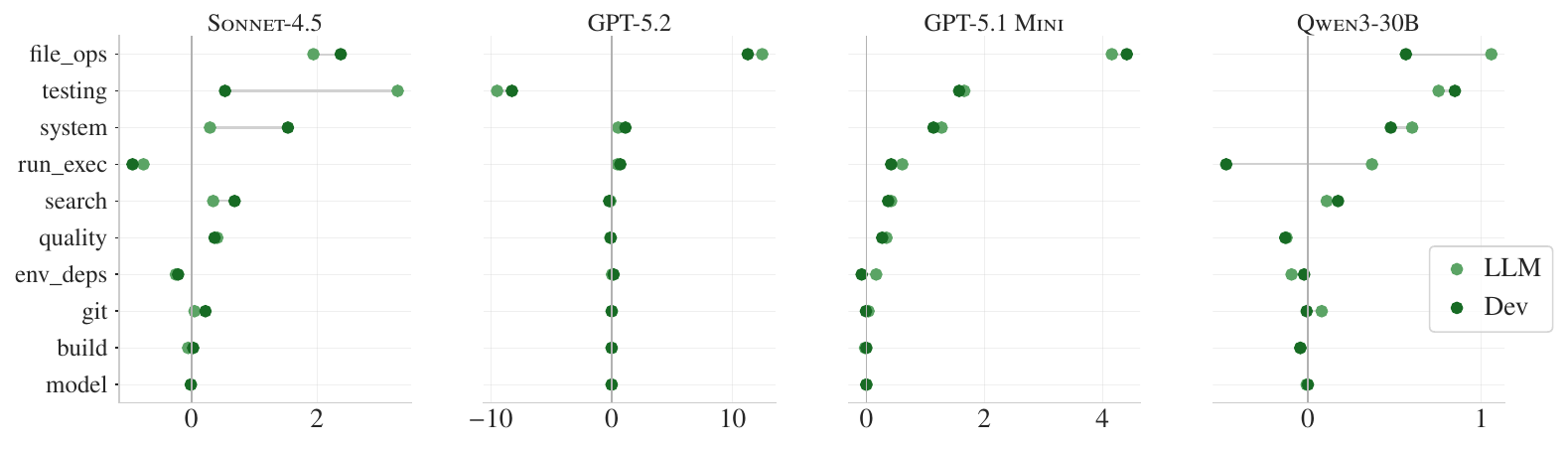}
    \caption{Increase in the average tool use (grouped into high-level categories) when including LLM-generated (bright green) or developer-provided (dark green) \agentsmds{}, compared to the average tool use without \agentsmds{}. For the high-level categories, we use an LLM to categorize the various tool calls.}
    \label{fig:experiment1_toolintent_freq}
\end{figure*}
\vspace{-0.1in}

\paragraph{Analyzing intents of tool calls}
For the tool intents extracted with the LLM judge ($334$ different categories in total), we further aggregate them into the following 10 categories:
\begin{itemize}
  \item \textbf{git}: Repository and version-control operations (e.g., commits, branches, diffs, checkout, stash, status).
  \item \textbf{model}: Model lifecycle tasks such as downloading or loading models, inspecting configurations or parameters, and running inference.
  \item \textbf{env\_deps}: Python environment and dependency management (virtualenv or venv, installations, versions, lockfiles).
  \item \textbf{build}: Building, compiling, or packaging code and producing artifacts or distribution packages.
  \item \textbf{quality}: Code quality and correctness checks (linting, formatting, type checking, validation or verification, schema checks).
  \item \textbf{testing}: Running and reviewing tests (unit, integration, regression, sanity, pytest) and test results.
  \item \textbf{run\_exec}: Executing workflows and scripts or commands (Python, shell, Django), including reproduction and debugging runs.
  \item \textbf{search}: Discovery and inspection actions (search, find, grep, glob, list, view, show, display, inspect, parse).
  \item \textbf{file\_ops}: Direct file and filesystem operations (read, write, edit, copy, move, delete, create, permissions, paths).
  \item \textbf{system}: System and miscellaneous utilities (processes, disk usage, environment variables, HTTP checks, checksums, tool or device information, help).
\end{itemize}
In \cref{fig:experiment1_toolintent_freq}, we show the difference in frequency of these categories with and without \agentsmds{}.
The conclusion is similar to that of \cref{sec:eval:trace}: the presence of \agentsmds{} significantly increases the number of tests run by coding agents, as well as the extent of codebase exploration and code quality checks.
\WFclear
\vspace{-0.1in}

\subsection{Per-repository Success Rate}

In \cref{fig:experiment1_perf_pre_repo}, we show the success rate of the different scenarios (\noplan{}, \autoplan{}, and \humanplan{}) grouped by repository.
For both \swebench{} and \planbench{}, there is no single repository where the presence of \agentsmds{} has a significant impact.
Nonetheless, for \planbench{} in particular, we see that the difficulty across instances is relatively balanced, validating our approach to building the instances.
\vspace{-0.1in}

\subsection{Additional Data}

\begin{table}[t]
    \caption{The success rate (in \%) plus or minus the standard error per \swebench{} and \planbench{}. We \textbf{bold} the best setting.}
    \label{tab:main_success}
    \centering
    \resizebox{0.7\linewidth}{!}{%
    \begin{tabular}{ll c c c c}
        \toprule
        & \multirow{1}{*}{Type} & \textsc{Sonnet-4.5} & \textsc{GPT-5.2} & \textsc{GPT-5.1 M.} & \textsc{Qwen3-30B} \\
        \midrule
        \multirow[c]{2}{*}{\shortstack[c]{\textsc{SWE-}\\\textsc{Bench}}} & None & $\mathbf{59.9 \pm 3.0}$ & $\mathbf{56.6 \pm 2.9}$ & $\mathbf{47.7 \pm 3.1}$ & $31.1 \pm 2.7$ \\
        \cmidrule{2-6}
        & LLM & $58.7 \pm 2.9$ & $54.4 \pm 2.9$ & $47.6 \pm 3.2$ & $\mathbf{32.4 \pm 2.7}$ \\
        \midrule
        \multirow[c]{3}{*}{\shortstack[c]{\textsc{Plan-}\\\textsc{Bench}}} & None & $\mathbf{73.2 \pm 3.8}$ & $65.2 \pm 4.1$ & $54.3 \pm 4.3$ & $45.7 \pm 4.3$ \\
        \cmidrule{2-6}
        & LLM & $65.2 \pm 4.1$ & $\mathbf{68.1 \pm 4.0}$ & $50.7 \pm 4.3$ & $47.1 \pm 4.3$ \\
        \cmidrule{2-6}
        & \textsc{Dev.} & $70.3 \pm 3.9$ & $\mathbf{68.1 \pm 4.0}$ & $\mathbf{55.8 \pm 4.2}$ & $\mathbf{53.6 \pm 4.3}$ \\
        \bottomrule
    \end{tabular}%
    }
\end{table}

In \cref{tab:main_success}, we report the success rate depicted in \cref{fig:experiment1_perf}.
\vspace{-0.1in}

\subsection{Additional Statistical Testing}

\begin{table}[!htb]
    \caption{Stratified permutation tests comparing the effect of different context files on cost and number of steps. Under the null hypothesis, we assume that the context files do not influence cost (respectively, the number of steps).}
    \label{tab:cost_pvalue}
    \centering
    \small
    \begin{tabular}{l l l l c c}
        \toprule
        Experiment & Benchmark & Metric & Comparison & p-value & Significant \\
        \midrule
        Experiment 1 & \swebench{}  & Steps & None vs LLM   & 0.0287   & Yes \\
        Experiment 1 & \swebench{}  & Cost  & None vs LLM   & $< 0.00001$ & Yes \\
        Experiment 1 & \planbench{} & Steps & None vs LLM   & 0.00004  & Yes \\
        Experiment 1 & \planbench{} & Steps & None vs Human & 0.00002  & Yes \\
        Experiment 1 & \planbench{} & Cost  & None vs LLM   & 0.00064  & Yes \\
        Experiment 1 & \planbench{} & Cost  & None vs Human & 0.0126   & Yes \\
        \bottomrule
    \end{tabular}
\end{table}

In \cref{tab:cost_pvalue}, we show the results of stratified permutation tests assessing the statistical significance of the increase in steps and costs induced by \agentsmds{}.
We find that \agentsmds{} systematically and significantly increase the average number of steps and the cost required to solve instances.

\section{Additional Results}
\label{app:ablations}

In this section, we propose an additional evaluation of the effects of \agentsmds{} on coding agents. 
In particular, we show that \agentsmds{} can act as effective overviews when no documentation is present. We evaluate how the length of \agentsmds{} influences our results, which categories within \agentsmds{} have the most impact, and whether contamination explains why we do not observe performance improvements when using \agentsmds{}.

\begin{wrapfigure}{r}{0.5\textwidth}
    \centering
    \includegraphics[width=.8\linewidth]{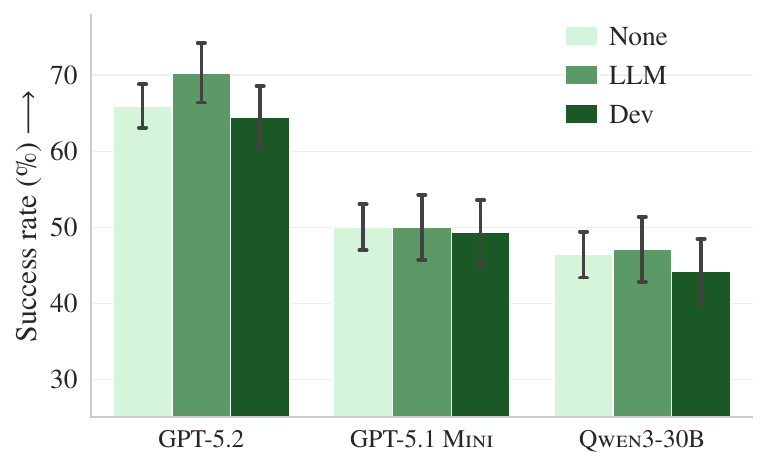}
    \caption{When removing all documentation-related files from the codebase, LLM-generated \agentsmds{} tend to outperform developer-provided (Human) ones on \colorbox{planbenchcolor}{\planbench{}}.
    }
    \label{fig:experiment4_perf}
\end{wrapfigure}

\paragraph{\Agentsmds{} are redundant documentation}
We show in \cref{sec:eval:main} that \agentsmds{} do not provide an effective repository overview.
Our hypothesis is that LLM-generated \agentsmds{} are mostly redundant with existing documentation, while developer-provided \agentsmds{} add additional information.
We confirm this hypothesis by manually removing all documentation (files ending with \texttt{.md}, example code, and the contents of the \texttt{docs/}) after generating the \agentsmd{} and before evaluating the coding agents, excluding \claudecode{} for cost reasons, and show the results in \cref{fig:experiment4_perf}.
In this setting, where \agentsmds{} are the only source of documentation available, we find that LLM-generated \agentsmds{} not only consistently improve performance by $2.7$\% on average, but also outperform developer-written ones across settings.
This may also explain anecdotal evidence reporting that coding agents perform better after adding \agentsmds{}~\citep{builderioImproveYourAICodeOutputWithAgentsMd}, since many less popular repositories contain little to no documentation.

\begin{figure*}[t]
    \centering
    \hspace*{\fill}
    \includegraphics[width=\linewidth]{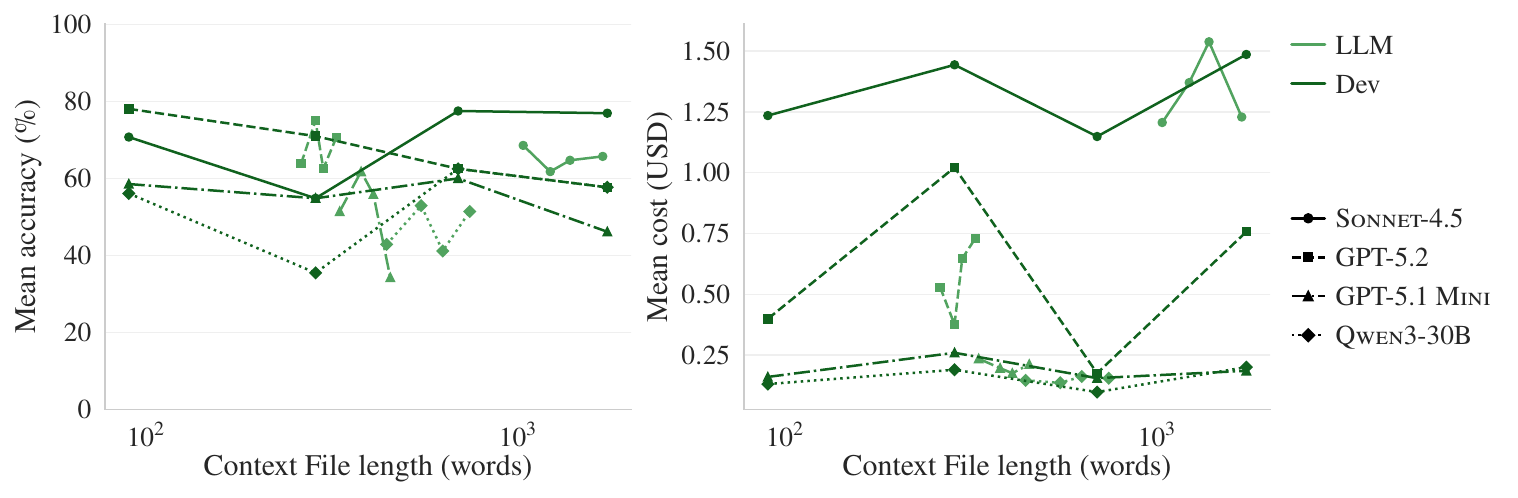}
    \hspace*{\fill}
    \hspace*{\fill}
    \caption{Resolution rate and average cost (in USD) for four different models with LLM-generated context files and with developer-provided ones on \planbench{}, binned by the length of the context files. We find no correlation between the resolution rate and the length of the context files.}
    \label{fig:context_file_length_exp1}
\end{figure*}

\begin{table}[t]
    \caption{Ablating common categories from LLM-generated \agentsmds{} does not significantly improve accuracy on either benchmark. For accuracy, p-values are computed with McNemar's test against the full \agentsmd{} condition; for cost, p-values are computed with a permutation test.}
    \label{tab:category_ablation}
    \centering
    \resizebox{\linewidth}{!}{%
    \begin{tabular}{llcccc}
        \toprule
        Benchmark & Metric & Full & \makecell{Without\\testing} & \makecell{Without\\overview} & \makecell{Without\\tooling} \\
        \midrule
        \multirow{2}{*}{\planbench{}} & Accuracy & $68.12\%$ & $66.67\%$ ($p=0.80$) & $62.32\%$ ($p=0.15$) & $63.77\%$ ($p=0.31$) \\
        & Cost & $\$0.4715$ & $\$0.3730$ (\textbf{$p=0.023$}) & $\$0.4027$ ($p=0.24$) & $\$0.4815$ ($p=0.97$) \\
        \midrule
        \multirow{2}{*}{\swebench{}} & Accuracy & $54.36\%$ & $57.72\%$ ($p=0.099$) & $54.20\%$ ($p=0.73$) & $53.69\%$ ($p=0.85$) \\
        & Cost & $\$0.3272$ & $\$0.2756$ (\textbf{$p=0.0035$}) & $\$0.3018$ ($p=0.10$) & $\$0.2715$ (\textbf{$p=0.0012$}) \\
        \bottomrule
    \end{tabular}%
    }
\end{table}

\paragraph{The length of \agentsmds{} does not influence our findings}
In \cref{fig:context_file_length_exp1}, we bin the success rate and per-instance cost results from \cref{fig:experiment1_perf} and \cref{tab:main_cost} by the length of the \agentsmd{}.
We observe no clear dependency between the success rate or the per-instance cost and the \agentsmd{} length. 
This suggests that the length of \agentsmds{} does not influence our findings, and that the increase in cost is better explained by the fact that coding agents tend to follow instructions written in the \agentsmds{} (see \cref{sec:eval:trace}).

\paragraph{No specific categories in LLM-generated \agentsmds{} help significantly}
We manually inspect LLM-generated \agentsmds{} and identify three categories that appear in most of them: \emph{Overview} (Does the context file explain the codebase structure), \emph{Tooling} (How to set up the environment? Which tools to use?), and \emph{Testing} (How to run tests in the codebase?).
For developer-provided \agentsmds{}, there are no clear categories shared between them.

To assess the importance of each category, we use \textsc{GPT-5.4} to remove each individual category from the \agentsmd{}, and then rerun \swebench{} and \planbench{} on \gptfive{}.
\cref{tab:category_ablation} shows that the testing category leads to a significant increase in cost in both \swebench{} and \planbench{}, and the tooling category leads to a significant increase in cost for \swebench{}.
Meanwhile, no category has a significant positive or negative effect on benchmark accuracy. 
Even if tendencies appear, such that omitting test harness details in \swebench{} increases performance, this is contradicted by a slightly negative effect of omitting it on \planbench{}.
Overall, these results back our recommendation that LLM-generated context files do influence behaviors (as seen by the cost increase) but ultimately do not significantly help improve coding agents' accuracy, even when considering only particular parts of the context file.

\begin{figure}[t]
    \centering
    \includegraphics[width=\linewidth]{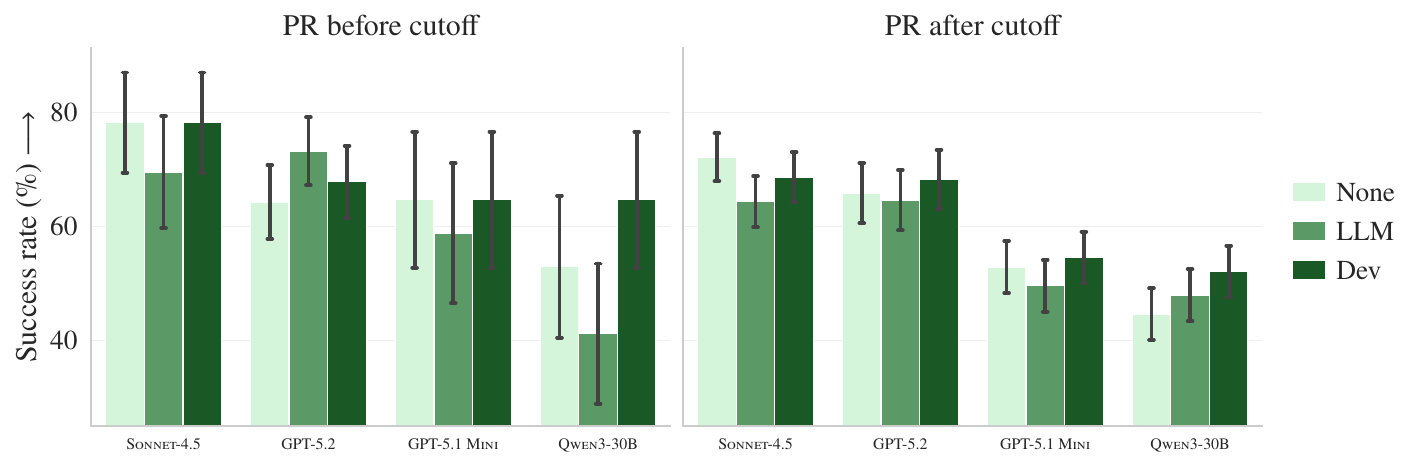}
    \caption{Resolution rate for four different models with LLM-generated context files and with developer-provided ones on \planbench{}, split by knowledge cutoffs.}
    \label{fig:contamination_ablation}
\end{figure}

\paragraph{Contamination does not influence our results}
To isolate the effect of contamination on our results, we split \planbench{} instances by whether the PR predates each model’s knowledge cutoff and compute accuracy for all three settings: no context files, LLM-generated context files, and human context files. 
Before the knowledge cutoff, this corresponds to $17$\% of instances for \sonnet{}, $40$\% of instances for \gptfive{}, and $12$\% of instances for \gptfivemini{} and \qwen{}. 
\cref{fig:contamination_ablation} shows no significant or consistent difference between pre- and post-cutoff instance success rates, suggesting that contamination does not affect our conclusions.

\section{Validating \planbench{} instances}
\label{app:analyzing_agentsmd}

In this section, we validate the quality of \planbench{}.
Specifically, we verify that developer-provided \agentsmds{} are significantly different from the LLM-generated ones.
We also manually inspect \planbench{} instances and validate the instance-description generation pipeline from \planbench{} by applying it to \swebench{}.

\begin{figure*}[t]
    \centering
    \hspace*{\fill}
    \includegraphics[width=.48\linewidth]{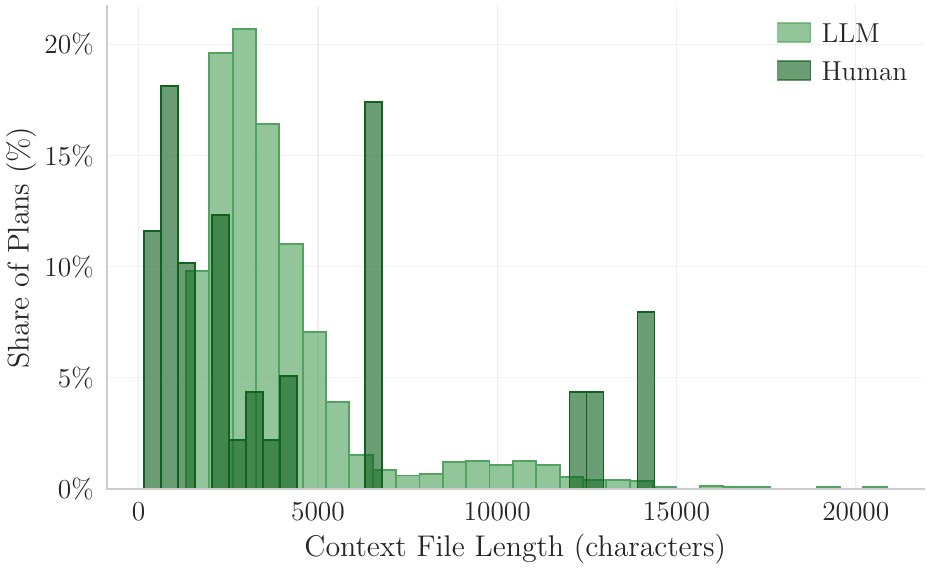}
    \hspace*{\fill}
    \includegraphics[width=.48\linewidth]{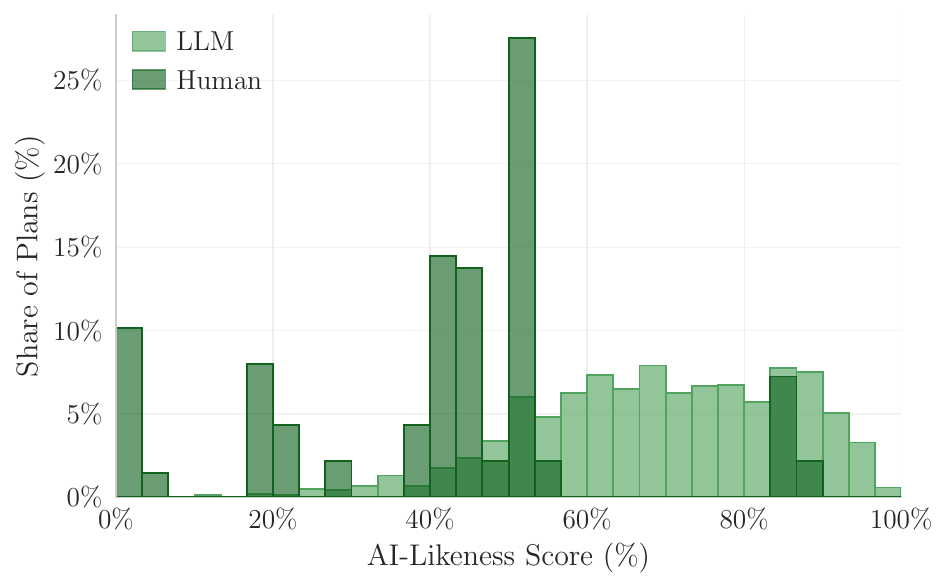}
    \hspace*{\fill}
    \caption{Length (left) and AI-detection score (right) distributions for LLM-generated and developer-provided \agentsmds{} on \planbench{}.
    }
    \label{fig:context_file_distrib}
\end{figure*}

\paragraph{Developer-provided \agentsmds{} differ from LLM-generated ones}
To compare developer-provided \agentsmds{} with LLM-generated ones, in \cref{fig:context_file_distrib}, we compare both their length distributions and their AI-detection scores.
For the AI-detection score, we use the open-source Pangram LLM detection model \textsc{EditLens Llama-3.2 3B}~\citep{pangram}.
If developer-provided \agentsmds{} were fully LLM-generated, they would follow distributions (both in length and AI-detection score) similar to those of the LLM-generated ones. 
Yet, we find in \cref{fig:context_file_distrib} that developer-provided \agentsmds{} tend to be shorter than LLM-generated ones and, according to the Pangram detector, are significantly less likely to be AI-generated.

\begin{figure}[t]
\begin{schemabox}
--- tests/pr/test_cli_lib.py (before)
+++ tests/pr/test_cli_lib.py (after)
@@ -45,7 +45,6 @@
 proc = run_cli("eval", "lib", str(prelude_path))

 assert proc.returncode != 0, "Evaluating a library should fail"
- assert "Can not evaluate a library" in proc.stderr


 def test_compile_lib_rejects_arguments() -> None:
@@ -53,4 +52,3 @@
 proc = run_cli("compile", "lib", str(prelude_path), "{}")

 assert proc.returncode != 0, "Supplying parameters to a library should fail"
- assert "Can not pass arguments to a library" in proc.stderr
\end{schemabox}
\caption{Example manual edit to a generated \planbench{} test for \texttt{opshin/opshin\#171}.}
\label{fig:overspecified_test_modification}
\end{figure}

\begin{table}[t]
        \caption{Resolution rate on \swebench{} for different models with either LLM-generated or original issue descriptions, under two context-file settings: no context files (None) and LLM-generated context files (LLM). Values in parentheses indicate the change relative to the corresponding no-context setting.}
    \label{tab:swebench_resolution_merged}
    \centering
    \small
    \begin{tabularx}{\linewidth}{l
        >{\centering\arraybackslash}X
        >{\centering\arraybackslash}X
        >{\centering\arraybackslash}X
        >{\centering\arraybackslash}X}
        \toprule
        & \multicolumn{2}{c}{No Context Files} & \multicolumn{2}{c}{LLM Context Files} \\
        \cmidrule(lr){2-3} \cmidrule(lr){4-5}
        Model & LLM-generated Desc. & Original Desc. & LLM-generated Desc. & Original Desc. \\
        \midrule
        \textsc{GPT-5.2}     & 70.8\% & 57.0\% & 66.8\% (-4\%)  & 54.9\% (-2\%)  \\
        \textsc{Qwen3-30B}   & 46.9\% & 30.3\% & 44.9\% (-2\%)  & 32.4\% (+2\%)  \\
        \bottomrule
    \end{tabularx}
\end{table}

\paragraph{\planbench{} instances are better specified than \swebench{}}
We manually inspect all \planbench{} instances by examining the task descriptions and the corresponding unit tests.
We find that most instances are well specified, with clear instructions on how to reproduce the issue and a technical description of the issue.
For 25 of the 138 instances, we manually edit the test cases to remove over-specified tests.
We provide an example of such a modification in \cref{fig:overspecified_test_modification}.
In this case, the test still verifies that evaluating or parameterizing a library fails, but it no longer requires a particular error message string.

To confirm our inspection, we further generate a variant of \swebench{} in which we generate problem descriptions using the \planbench{} pipeline (see \cref{sec:method}), which enables us to evaluate the concrete impact of LLM-generated task descriptions compared to human baselines.
We observe in \cref{tab:swebench_resolution_merged} a consistent trend of about 15\% higher accuracy, preserving all model rankings (with and without context files).
We thus conclude that the \planbench{} construction results in a slightly easier but fairer evaluation for coding agents compared to other benchmarks such as \swebench{}.

\section{Asset Licenses}
\label{app:licenses}

In this section, we list the licenses of all assets used in this work.

\paragraph{Datasets}
\begin{itemize}
    \item \swebench{} \textsc{Lite}: MIT.
    \item \planbench{}: all source repositories used to build the dataset have explicit open-source licenses that permit our use; the licenses are Apache-2.0, MIT, AGPL-3.0, GPL-3.0, and BSD-3-Clause.
\end{itemize}

\paragraph{Models and coding agents}
\begin{itemize}
    \item \sonnet{} and \textsc{Haiku-4.5}: Anthropic Commercial Terms of Service.
    \item \gptfive{} and \gptfivemini{}: OpenAI Services Terms.
    \item \qwen{}: Apache-2.0.
    \item \gptoss{}: Apache-2.0.
    \item \codex{}: Apache-2.0.
    \item \qwencode{}: Apache-2.0.
    \item \claudecode{}: Anthropic Commercial Terms of Service.
\end{itemize}

\section{Prompts}
\label{app:prompt}

In this section, we detail all prompts used throughout this work.
\vspace{-0.1in}

\subsection{\planbench{} instances generation}

We detail below the prompts used for filtering pull requests, setting up the instances, describing the instances, and generating the test cases.

\begin{promptbox}[Filtering pull requests]
You are evaluating pull request \placeholder{pr\_number} for suitability as a regression-test
task in SWE-bench style datasets. Decide whether the PR primarily introduces deterministic, testable behaviour.
Such behaviors typically include bug fixes, but can also include feature additions as long as it is possible to write a precise specification that allows testing the new feature independently of the implementation.

\medskip
\textbf{Repository:} \placeholder{repo\_full\_name}\\
\textbf{Title:} \placeholder{title}\\
\textbf{Author:} \placeholder{author}\\
\textbf{Merged at:} \placeholder{merged\_at}

\medskip
\textbf{PR description}\\
\placeholder{body}

\medskip
\textbf{Diff excerpt}\\
\placeholder{excerpt}

\medskip
\textbf{Deliverables}
\begin{enumerate}[leftmargin=1.6em, itemsep=0.25em]
  \item Do \textbf{not} modify existing project code.
  \item Create the JSON file \placeholder{decision\_path} with UTF-8 encoded content describing your
  decision using this schema:

  \begin{schemabox}
{
  "pr_number": <int>,
  "suitable": <bool>,
  "needs_manual_review": <bool>,
  "decision": "include" | "exclude" | "manual_review",
  "rationale": "<short explanation>",
  "key_files": ["relative/file.py", "..."],
  "risk_factors": ["<short string>", "..."]
}
  \end{schemabox}

  \begin{itemize}[leftmargin=1.6em, itemsep=0.25em]
    \item Set \texttt{"decision"} to \texttt{"include"} only when you are confident the PR is a self-contained
    bug fix that can be validated via regression tests.
    \item Use \texttt{"manual\_review"} if you are uncertain.
  \end{itemize}
  \item Stage the JSON file and finish. Do not stage anything else.
\end{enumerate}
\end{promptbox}

\begin{promptbox}[Setting up the instance]
Your goal is to help developers set up their environment to run code in the repository and be able to run the current tests. You should write a list of all commands needed to (i) set up the environment from scratch, and (ii) run the existing tests. You need to make sure that the commands you provide actually work for you. The setup is considered valid if most of the tests are passing after running \textbf{exactly} your setup commands and the test commands you provide.

\medskip
\textbf{Test runner requirement}\\
To run the repository tests, create a file at the root of the repository called \texttt{run\_tests.py} that:
\begin{itemize}[leftmargin=1.6em, itemsep=0.25em]
  \item executes all tests,
  \item parses the test output,
  \item writes a JSON file at the repository root named \texttt{test\_results.json} with schema:
\end{itemize}

\begin{schemabox}
{"test_name": <bool>, ...}
\end{schemabox}

where each \texttt{test\_name} is the name of a test and the boolean indicates whether the test passed (\texttt{true}) or failed (\texttt{false}).

\medskip
\textbf{Deliverables}
\begin{enumerate}[leftmargin=1.6em, itemsep=0.25em]
  \item Create the JSON file \placeholder{decision\_path} with UTF-8 encoded content explaining the steps to set up the environment and run the tests (using the \texttt{run\_tests.py} script you created):
  \begin{schemabox}
{
  "setup_commands": ["<command1>", "<command2>", "..."],
  "test_commands": ["<command1>", "<command2>", "..."]
}
  \end{schemabox}
  \item Create the script \texttt{run\_tests.py} at the root of the repository.
  \item Stage the JSON file and the script and finish. Do not stage anything else.
\end{enumerate}

\medskip
\placeholder{example\_files\_section}
\end{promptbox}

\begin{promptbox}[Describing the instance]
You are given a pull request (PR) and the related issues for a given GitHub repository.
Your goal is to format this information into a clear GitHub Issue following the template below.

\begin{itemize}[leftmargin=1.6em, itemsep=0.25em]
  \item For the \textbf{Steps to Reproduce} field, only write the steps you actually took to reproduce the issue in your specific environment. Make those steps \textbf{reproducible and minimal}.
  \item Developers should be able to implement a solution similar to the one provided in the PR, but the Issue should \textbf{not leak the solution}.
  \item Save your output in Markdown format in the file \placeholder{metadata\_relpath}.
\end{itemize}

\medskip
\textbf{Feature requests: Specification required}\\
Additionally, for issues about \textbf{adding a new feature} (rather than fixing a bug), include a \textbf{precise Specification} describing the desired behavior. It must be detailed enough to allow independent testing without relying on implementation details from the PR.

\begin{itemize}[leftmargin=1.6em, itemsep=0.25em]
  \item Specify inputs (types, valid ranges, edge cases), outputs, side effects, and any required error handling.
  \item If the PR includes human-readable outputs (logs, UI text, error messages, \dots), include them in the specification and state that fixes must use \textbf{exactly} those messages.
\end{itemize}

\medskip
\textbf{Issue template (copy into your Markdown output)}
\begin{schemabox}
### Description
(Provide a clear and concise description of the problem.)

### Steps to Reproduce
1. [Step 1]
2. [Step 2]
3. ...

### Expected Behavior (if applicable)
(Explain what you expected to happen.)

### Actual Behavior (if applicable)
(Explain what actually happened.)

### Specification (if applicable)
(Provide a precise specification of the desired behavior.)

### Additional Information
(Add screenshots, logs, or other helpful details.)
\end{schemabox}

\medskip
\textbf{Data for PR \#}\placeholder{pr\_number}\textbf{ in repository }\placeholder{repo}\textbf{ at commit }\placeholder{commit\_sha}

\medskip
\textbf{PR description}\\
\placeholder{pr\_description}

\medskip
\textbf{Referenced issues mentioned in the PR}\\
\placeholder{referenced\_issues\_text}

\medskip
\textbf{PR patch}\\
\placeholder{pr\_patch}

\medskip
\textbf{PR test (if any)}\\
\placeholder{pr\_test\_patch}

\medskip
\textbf{Key files identified during triage}\\
\placeholder{key\_files\_text}
\end{promptbox}

\begin{promptbox}[Generating the test cases]
You are generating regression tests for pull request \placeholder{pr\_number} in \placeholder{repo}. The current checkout is the base (pre-fix) commit \placeholder{commit\_sha}.

\medskip
\textbf{Problem description}\\
\placeholder{problem\_description}

\medskip
\textbf{PR patch}\\
\placeholder{pr\_patch}

\medskip
\textbf{PR test (if any)}\\
\placeholder{pr\_test\_patch}

\medskip
\textbf{Requirements}
\begin{enumerate}[leftmargin=1.6em, itemsep=0.35em]
  \item Focus on \textbf{deterministic} tests that expose the bug fixed by this PR. Tests should target \textbf{expected behavior} and must \textbf{not} rely on internal implementation details (variables, hidden helpers, etc.). They should fail on the base commit and pass on the merge commit (after applying the PR patch). You must verify this property. You may apply the provided patch using \texttt{git apply}. If a specification is provided in the problem description, tests must \textbf{exactly} align with it. Avoid tests that depend on incidental choices (variable names, function names, strings, \dots) unless explicitly required by the specification.
  \item Create \texttt{run\_pr\_tests.py} at the repository root that executes \emph{only} the tests you created, parses test output, and writes JSON results to \texttt{pr\_test\_results.json} with schema:
  \begin{schemabox}
{"test_name": <bool>, ...}
  \end{schemabox}
  You may use \texttt{run\_tests.py} as a reference. Note: your script should only run the tests you created for this PR.
  \item Ensure new tests match the project's existing test style and conventions. First review existing tests to understand structure and framework. You may reuse tests from the PR if appropriate.
  \item All new tests must be in \textbf{new files} created as part of this work. Do \textbf{not} modify any existing test files.
  \item For \texttt{test\_commands}, include any necessary steps (sourcing environments, setting variables, etc.) so tests run correctly in a fresh shell.
\end{enumerate}

\medskip
\textbf{Deliverables}
\begin{enumerate}[leftmargin=1.6em, itemsep=0.35em]
  \item Create the new test files with your proposed tests.
  \item Create the JSON file \placeholder{metadata\_relpath} with UTF-8 content explaining how to run the tests:
  \begin{schemabox}
{
  "test_commands": ["<command1>", "<command2>", "..."],  # Commands to run the PR tests with `run_pr_tests.py`
  "test_files": ["path/to/test_file1", "path/to/test_file2", "..."]
}
  \end{schemabox}
  \item Create the script \texttt{run\_pr\_tests.py} at the root of the repository.
  \item Stage the JSON file and the script and finish. Do not stage anything else.
\end{enumerate}
\end{promptbox}

\subsection{Analyzing Traces of Coding Agents}
\label{app:prompts:trace_analysis}

To analyze the tool calls made by the coding agents, we use \gptoss with the prompt below.

\begin{promptbox}[Analyzing coding agent traces]
You are labeling a tool call with a single intent category.

\medskip
\textbf{Goal:} choose a category name that is:
\begin{itemize}[leftmargin=1.6em, itemsep=0.25em]
  \item \textbf{Right-sized granularity:} more specific than ``execute command'' but not tied to exact arguments.
  \item \textbf{Reusable:} should apply to many future tool calls.
  \item \textbf{Clean:} do \textbf{not} include file paths, flags, quoted strings, IDs, repo names, or counts.
  \item \textbf{Format:} 2--5 words, lowercase, verb + object (e.g., \texttt{run tests}, \texttt{search codebase}). Avoid too-generic names like \texttt{run scripts}; specify what the script does (e.g., \texttt{compile code}).
\end{itemize}
You must also explain which tool is being used (e.g., \texttt{pytest}, \texttt{rg}, \dots) in a dedicated field.

\medskip
\textbf{You will be given}
\begin{itemize}[leftmargin=1.6em, itemsep=0.25em]
  \item \textbf{tool\_call:} the command or structured tool invocation
  \item \textbf{tool\_output:} optional output text
\end{itemize}

\medskip
\textbf{Existing categories (use one if it fits):} \placeholder{existing\_tool\_names}

\medskip
\textbf{Decision rules}
\begin{enumerate}[leftmargin=1.6em, itemsep=0.25em]
  \item If one existing category fits, use it \textbf{exactly}.
  \item If none fit, create \textbf{one} new category that:
  \begin{itemize}[leftmargin=1.6em, itemsep=0.2em]
    \item is \textbf{not} tool-specific (avoid \texttt{pytest}, \texttt{kubectl}, \texttt{terraform}, etc.)
    \item would likely match \textbf{5+} future tool calls
  \end{itemize}
  \item If the tool call does multiple things, pick the \textbf{primary intent} as the category (mention secondary intents in reasoning).
\end{enumerate}

\medskip
\textbf{Return JSON only}
\begin{schemabox}
{
  "tool_name": "<category>",
  "tool_used": "<specific tool or executable being invoked>",
  "reasoning": "<1-3 sentences: why this is the primary intent; include key clues from call/output; mention secondary intents if any>"
}
\end{schemabox}

\medskip
\textbf{Tool call}
\placeholder{tool\_call}

\medskip
\textbf{Tool output (if any)}
\placeholder{tool\_output}
\end{promptbox}

\clearpage

\relax

\end{document}